\documentclass[preprint]{jpsj3}
\usepackage{color}
\usepackage[T1]{fontenc}

\title{
Emergence of a Dimer--Dimer Interaction in the Low-Energy Effective Quantum-Dimer Model of a Diamond-Like-Decorated Square-Lattice Heisenberg Antiferromagnets with Further Neighbor Couplings
}
\author{Yuhei Hirose, Akihide Oguchi, and Yoshiyuki Fukumoto}
\inst{Department of Physics, Faculty of Science and Technology,\\ Tokyo University of Science, Noda, Chiba 278-8510, Japan}
\recdate{\today}
\abst{
We study spin-1/2 Heisenberg antiferromagnets on a diamond-like-decorated square lattice perturbed by two kinds of further neighbor couplings.
In our previous study [J. Phys. Soc. Jpn. {\bf85}, 094002 (2016)], the second-order effective Hamiltonian for the Heisenberg model perturbed by a further neighbor coupling was found to be a square-lattice quantum-dimer model with a finite hopping amplitude, $t>0$, and no dimer--dimer interaction, $v=0$.
In this study, we introduce another kind of further neighbor coupling and show that it leads to an attractive interaction between dimers, which suggests the stabilization of the columnar phase of the square-lattice quantum-dimer model.
The calculated $v/t$ is presented as a function of the ratio of the two exchange parameters in the Heisenberg model.
}

\begin{document}
\maketitle

\section{Introduction}\label{sec:1}

Since  Anderson proposed the resonating valence bond (RVB) theory for spin-1/2 two-dimensional Heisenberg antiferromagnets in 1973\cite{Anderson1973}, 
the exploration of RVB states has been one of the central issues in condensed matter physics.
In 1988, Rokhsar and Kivelson proposed the quantum dimer model (QDM)\cite{Rokhsar1988} as a phenomenological Hamiltonian of the RVB theory, motivated by the discovery of a high-temperature cuprate superconductor.
The QDM is written by 
\begin{equation}
\begin{minipage}[c]{.93\textwidth}
\begin{center}
\includegraphics[width=1.0\linewidth]{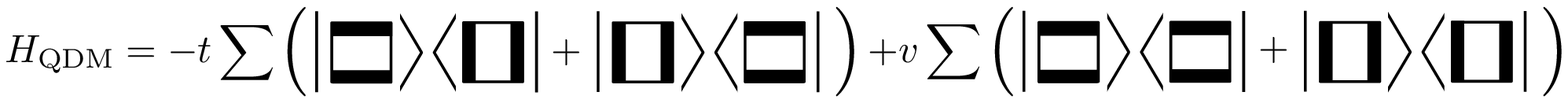}
\end{center}
\end{minipage}
\label{eq:1},
\end{equation}
where $t$ and $v$ represent the pair-hopping amplitude and dimer--dimer interaction, respectively. 
When the interaction is dominant, $|v|\gg t$, the staggered phase appears for $v/t\to \infty$ and the columnar phase appears for $v/t\to -\infty$.
Interestingly, at $v/t=1$, which is called the Rokhsar--Kivelson (RK) point, 
the ground state is the equal-amplitude superposition of dimer-covering configurations, which is ascribed to an RVB spin--liquid state. \cite{Rokhsar1988, square_qdm_1996, square_qdm_2006, square_qdm_2008, Moessner2008, book, square_qdm_2014, square_qdm_2016}
However, in the region of $v/t<1$, there has been substantial disagreement about the determination of the phases\cite{square_qdm_1996, square_qdm_2006, square_qdm_2008, Moessner2008, book, square_qdm_2014, square_qdm_2016}. Leung et al. have suggested that the columnar phase is stabilized in the region of $-\infty<v/t<-0.2$ and the plaquette phase appears for $-0.2<v/t$ by diagonalizing the Hamiltonian of the model on square lattices with periodic boundary conditions\cite{square_qdm_1996}. However, . Sylju\r{a}sen has argued that the columnar phase persists until $v/t\sim+0.6$ by using a continuous-time reptation quantum Monte Carlo method\cite{square_qdm_2006}. Furthermore, Ralko et al. have found evidence of a mixed columnar--plaquette phase around $v/t=0$ by using exact diagonalization and the Green's function Monte Carlo method\cite{square_qdm_2008}. 
Recently, Banerjee et al. have pointed out that the columnar phase extends all the way to $v/t=1$ without any plaquette or mixed phases by using exact diagonalization and a quantum Monte Carlo method\cite{square_qdm_2014,square_qdm_2016}. 
In this way, the determination of the phases of the square-lattice QDM in the region of $v/t<1$ is controversial.

The study of the QDM has been extended to various lattices, such as the triangular lattice\cite{Moessner2008,book,Moessner2001,Moessner2002,Moessner2005} and simple cubic lattice\cite{Moessner2008,book,Huse2003,Moessner2003,Hermele2004}.
For triangular and simple cubic lattices, it has been reported that the RVB liquid emerges in the finite area of $v/t\le 1$.\cite{Moessner2008,book,Moessner2001,Moessner2002,Moessner2005,Huse2003,Moessner2003,Hermele2004}
In this way, fascinating states such as RVB liquids appear in the QDM; however, it has not been made clear whether the QDM can be realized from realistic quantum spin Hamiltonians. 
Actually, a great deal of effort has been devoted to constructing QDMs from quantum spin Hamiltonians for the purpose of discovering RVB liquids. 
For example, Fujimoto has derived QDMs from two-dimensional antiferromagnetic quantum spin systems and found spin--liquid ground states in these systems\cite{Fujimoto}. 
However, the experimental realization of the QDMs in these systems may be difficult because the original spin models contain complicated multiple spin interactions.

The main stage for the study of quantum spin liquids is the spin-1/2 kagome-lattice Heisenberg antiferromagnet.
The ground state of this system has been theoretically argued to be a spin liquid,\cite{Yan2011,Depenbrock2012} 
but corresponding experimental systems have not been found yet, despite tremendous effort to synthesize kagome compounds.
For example, although herbertsmithite $\rm{ZnCu_3(OH)_6Cl_2}$ has been suggested to be an ideal kagome compound with a quantum spin-liquid state,\cite{herbertsmithite1,herbertsmithite2,herbertsmithite3}
Kawamura et al. have pointed out that the intrinsic randomness in herbertsmithite results in a gapless ``random singlet state''\cite{Kawamura1,Kawamura2}.

In 2006, Kitaev proposed a quadratic spin Hamiltonian with a spin--liquid ground state, 
which is now called the Kitaev model and provides another route for the exploration of spin liquids.\cite{Kitaev1, Kitaev2} 
The Kitaev model is defined on a honeycomb lattice and contains anisotropic ferromagnetic interactions.
Owing to the anisotropic ferromagnetic interactions, the Kitaev model is approximately realized in $\rm{Na}_2\rm{Ir}\rm{O}_3$, 
but this compound is known to have magnetic order\cite{Kitaev3}.

Needless to say, it would be a significant step toward the exploration of spin liquids to find a quadratic spin Hamiltonian that yields the QDM.
Recently, it has been proposed that the QDM can be realized as a low-energy effective Hamiltonian of a spin-1/2 Heisenberg antiferromagnet on a diamond-like-decorated square lattice\cite{Hirose2016}.
The spin Hamiltonian contains only quadratic Heisenberg-type exchange terms, which is very important for experimental realizations such as, for example, in a quantum simulator using optical lattices\cite{Bloch2008}. 
Furthermore, if our diamond-like-decorated square lattice is realized by optical lattices, a Raman scattering experiment can enable us to identify whether the square-lattice QDM is realized as the low-energy effective Hamiltonian. This is because the square-lattice QDM is the effective Hamiltonian for the low-energy singlet sector and the singlet sector can be observed in a Raman scattering experiment\cite{raman1}.

\begin{figure}[t] 
\begin{center}
\includegraphics[width=.55\linewidth]{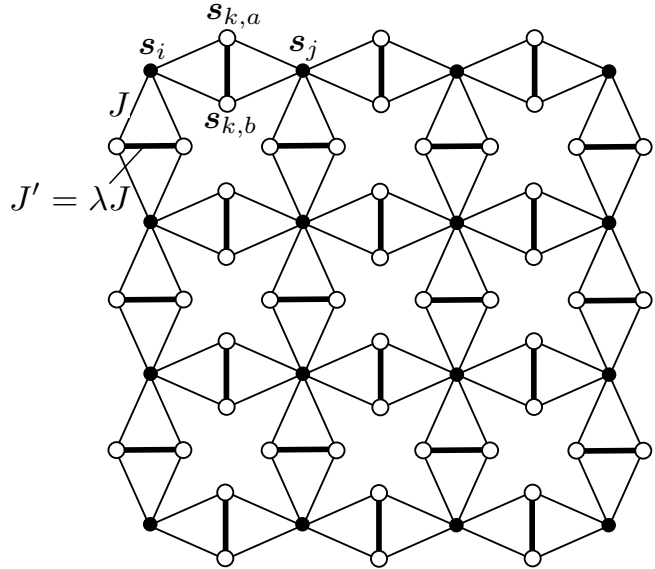}
\end{center}
\caption{Structure of a diamond-like-decorated square lattice. The thin and thick solid lines represent the antiferromagnetic interactions $J$ and $J'=\lambda J$, respectively. We call $\mib{s}_i$ and $\mib{s}_j$ the edge spins and the pair ($\mib{s}_{k,a}, \mib{s}_{k,b}$) a bond spin-pair. The edge spins and the bond spin-pairs are indicated by the closed and open circles, respectively, and the magnitude of the spin operators is $1/2$.  
 }
\label{fig:1}
\end{figure}

A diamond-like-decorated square lattice is a lattice in which the bonds of a square lattice are replaced with diamond units, as shown in Fig.~\ref{fig:1}.
For this lattice, if we define the interaction strength of the four sides of a diamond unit as $J$ and that of the diagonal bond as $J'=\lambda J$, 
the ratio $\lambda$ determines the ground-state properties\cite{Hirose2017}.
As shown in Fig.~\ref{fig:1}, we denote the four $S=1/2$ operators in a diamond unit as $\mib{s}_i$, $\mib{s}_j$, $\mib{s}_{k,a}$, and $\mib{s}_{k,b}$.
We call $\mib{s}_i$ and $\mib{s}_j$ edge spins (closed circles in Fig.~\ref{fig:1}) and the pair ($\mib{s}_{k,a}, \mib{s}_{k,b}$) a bond spin-pair (open circles). 
The ground state of an isolated diamond unit for $\lambda<2$ becomes a four-spin singlet state (tetramer singlet state), where both edge spins and the bond spin-pair are in triplet pair states.\cite{Hirose2017,Takano1996,Morita2016}
For a diamond-like-decorated square lattice with $0.974<\lambda<2$, the ground-state manifold consists of macroscopically degenerated tetramer-dimer (MDTD) states [see Fig.~3(a) in Ref.~30].
If we regard a tetramer as a ``dimer'' in the QDM, then MDTD states are equivalent to square-lattice dimer-covering states.\cite{Hirose2016,Hirose2017,Morita2016}

\begin{figure}[t]
\begin{center}
\includegraphics[width=.9\linewidth]{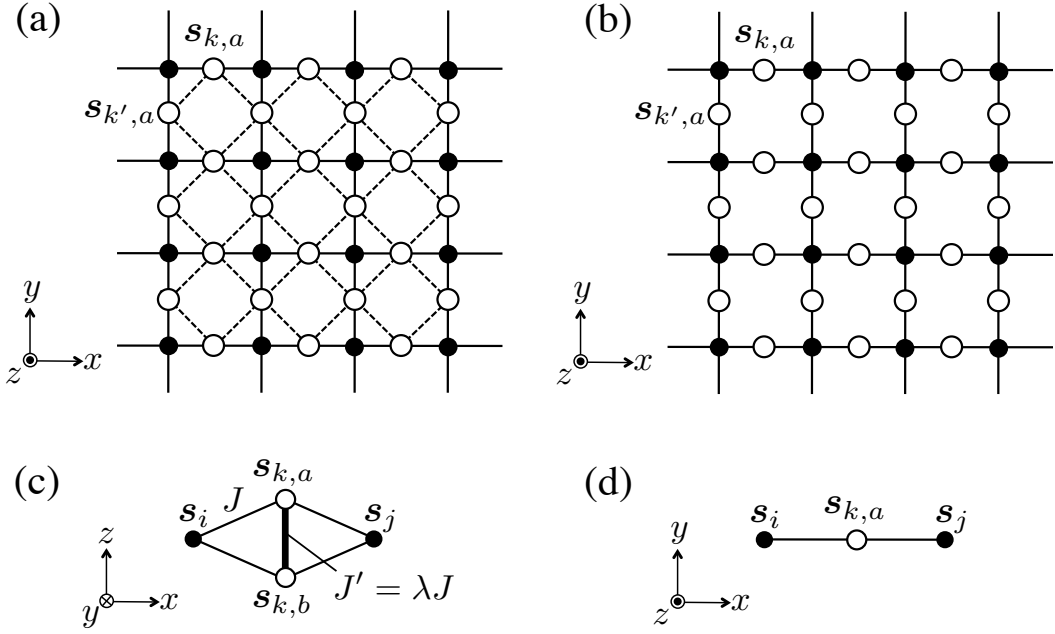}
\end{center}
\caption{(a) Structure of the diamond-like-decorated square lattice introducing the further neighbor couplings, where the direction of the bond spin-pairs is along the $z$-axis. Further neighbor couplings, which are indicated by the dashed lines, connect $\mib{s}_{k,a}$ and $\mib{s}_{k',a}$ and connect $\mib{s}_{k,b}$ and $\mib{s}_{k',b}$ between the orthogonally connected diamond units, although only the former is drawn. (b) Structure of the diamond-like-decorated square lattice before introducing the further neighbor couplings. (c) Top and (d) side views of the diamond unit.
}
\label{fig:2}
\end{figure}

In our previous study,\cite{Hirose2016}  we calculated the second-order effective Hamiltonian for a model with a further neighbor coupling, which is shown by the dashed lines in Fig.~\ref{fig:2}(a). 
In Fig.~\ref{fig:2}(b), we draw the diamond-like-decorated square lattice before introducing the further neighbor coupling, which is somewhat different from the lattice shown in Fig.~\ref{fig:1}. 
In the lattice in Fig.~\ref{fig:1}, we choose the condition that the direction of the bond spin-pairs is parallel to the plane formed by the edge spins. 
On the other hand, in the other lattice in Fig.~\ref{fig:2}(b), we choose the direction of the bond spin-pairs to be orthogonal to the edge-spin plane, i.e., 
we make the direction of the bond spin-pairs parallel to the $z$-axis, as shown in Figs. 2(c) and 2(d).
Therefore, in the geometry of the diamond-like-decorated square lattice of Figs.~\ref{fig:2}(a) and 2(b), we have three layers: 
the center layer composed of edge spins at $z=0$, the upper layer composed of $\{\mib{s}_{k,a}\}$ at $z>0$, and the lower layer composed of $\{\mib{s}_{k,b}\}$ at $z<0$.
The further neighbor coupling connects $\mib{s}_{k,a}$ and $\mib{s}_{k',a}$ and connects $\mib{s}_{k,b}$ and $\mib{s}_{k',b}$, 
where $k$ and $k'$ represent two adjacent diamond units in the same plaquette, and only the former is drawn in Fig.~\ref{fig:2}(a). 
For the model in Fig.~\ref{fig:2}(a), our obtained second-order effective Hamiltonian was a square-lattice QDM with a finite pair-hopping amplitude ($t>0$) and no dimer--dimer interaction ($v=0$).\cite{Hirose2016}
However, it is desirable to derive QDMs with $v\neq0$ because various phases appear, depending on the magnitude of $v/t$.\cite{square_qdm_1996, square_qdm_2006, square_qdm_2008, Moessner2008, book, square_qdm_2014, square_qdm_2016}
Because $v/t$ is a function of the parameter $\lambda$ in the original spin Hamiltonian, it is possible to obtain various phases in the QDM to appear by changing the value of $\lambda$.
For example, if the value of $\lambda$ is chosen so that $v/t=1$, this means that the equal-amplitude RVB liquid is found\cite{Rokhsar1988, square_qdm_1996, square_qdm_2006, square_qdm_2008, Moessner2008, book, square_qdm_2014, square_qdm_2016}.
Furthermore, because there have been various controversies concerning the determination of the phase diagram of the QDM on a square lattice, as mentioned above, if we find the value of $\lambda$ corresponding to the region of $v/t<1$, it is worth investigating what kind of phase appears in our spin system.

\begin{figure}[h]
\begin{center}
\includegraphics[width=.7\linewidth]{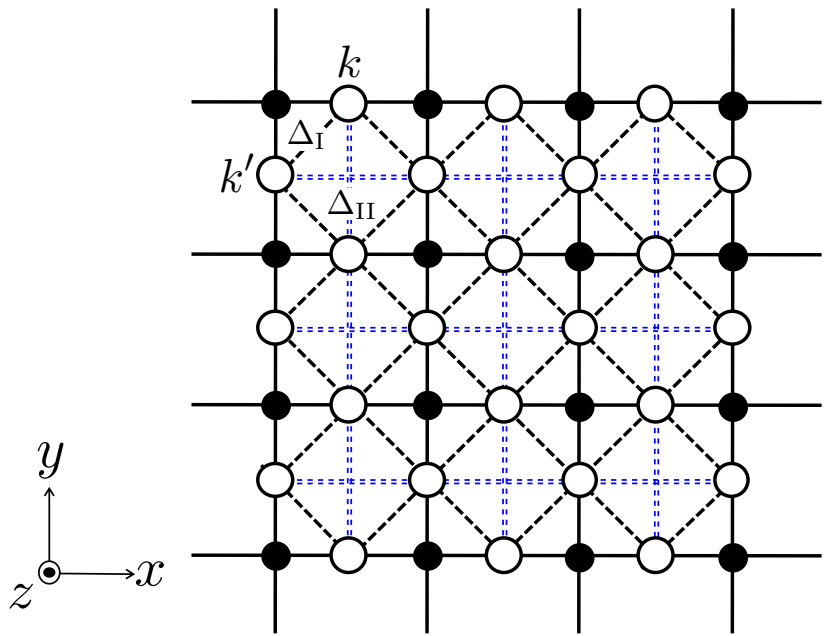}
\end{center}
\caption{(Color online) Structure of the diamond-like-decorated square lattice with the introduction of two kinds of further neighbor couplings $\Delta_{\rm{I}}$ and $\Delta_{\rm{II}}$. The further neighbor couplings $\Delta_{\rm{I}}$, which are indicated by the dashed lines, correspond to the dashed lines in Fig.~\ref{fig:2}(a).
The couplings $\Delta_{\rm{II}}$, which are indicated by the double dashed lines, connect the bond spins of the upper layer to each other and connect those of the lower layer to each other, as is the case with the couplings $\Delta_{\rm{I}}$.}
\label{fig:3}
\end{figure}

In this study, we introduce two kinds of further neighbor couplings, as shown in Fig.~\ref{fig:3}, and calculate the second-order effective Hamiltonian for the model.
The further neighbor coupling $\Delta_{\rm{I}}$, which is shown by the dashed lines in Fig.~\ref{fig:3}, corresponds to the dashed lines in Fig.~\ref{fig:2}(a).
In our previous study, we only introduced the coupling $\Delta_{\rm{I}}$, but in the present study, we consider an additional coupling $\Delta_{\rm{II}}$ between two diamond units facing each other in the same plaquette, which is shown by the double dashed lines in Fig.~\ref{fig:3}.
Then, we explain the reason why we introduce the couplings $\Delta_{\rm{II}}$.
Taking into consideration the fact that the magnitude of couplings depends on the orbital overlap, 
it is natural to think that the couplings $\Delta_{\rm{II}}$ exist if there are couplings $\Delta_{\rm{I}}$ and the magnitude of couplings $\Delta_{\rm{II}}$ is  comparable to that of $\Delta_{\rm{I}}$.
As is the case with the coupling $\Delta_{\rm{I}}$, the coupling $\Delta_{\rm{II}}$ connects the bond spins in the upper layer to each other and those in the lower layer to each other. 
When we introduce the coupling $\Delta_{\rm{I}}$ only, we obtain $v=0$ because all the contributions of perturbation processes cancel out with each other.\cite{Hirose2016}
On the other hand, this is not the case for the coupling $\Delta_{\rm{II}}$, which leads to $v\neq0$, as will be shown later. 
As a result, we obtain a square-lattice QDM with $-1.2\le v/t\le0$, i.e., a finite pair-hopping amplitude ($t>0$) and a negative dimer--dimer interaction ($v<0$). Here, we explain which phases in the phase diagram of the QDM on a square lattice correspond to our obtained phases.
If we use the result of Leung et al., who pointed out that the phase boundary between the plaquette and columnar phases is $v/t=-0.2$\cite{square_qdm_1996}, our obtained phase corresponds to the plaquette and columnar phases, which involve the phase boundary between these phases. In our previous study\cite{Hirose2016}, we suggested that our obtained phase with $v=0$ and $t>0$ results in the plaquette phase based on the phase diagram of Leung et al.\cite{square_qdm_1996}
However, in recent studies\cite{square_qdm_2006, square_qdm_2008, square_qdm_2014, square_qdm_2016, Moessner2008, book}, it has been shown that the columnar phase is stabilized in the region of $v/t\le0$. Therefore, by using the result of the recent studies\cite{square_qdm_2006, square_qdm_2008, square_qdm_2014, square_qdm_2016, Moessner2008, book}, we suggest that our obtained result with $-1.2\le v/t\le0$ corresponds to the columnar phase.

This paper is organized as follows.
The Heisenberg model on a diamond-like-decorated square lattice and the perturbation Hamiltonian are defined in Sect. 2.
In Sect. 3, we derive a square-lattice QDM as a second-order effective Hamiltonian and describe the details of the perturbation calculation process.
In Sect. 4, we summarize the results obtained in this study.

\section{Model}\label{sec:2}

\subsection{Diamond unit and tetramer ground state}

For the diamond unit in Fig.~\ref{fig:2}, we define
\begin{equation}
   h_{i,j}=(\mib{s}_i+\mib{s}_j)\cdot(\mib{s}_{k,a}+\mib{s}_{k,b})+\lambda \left(\mib{s}_{k,a}\cdot\mib{s}_{k,b}+\frac{3}{4}\right),
   \label{eq:2} 
\end{equation}
where we assume $J=1$ for simplicity. 
In the case of $\lambda<2$, the lowest eigenvalue of $h_{i,j}$ is $\lambda-2$
and the corresponding eigenvector $|\phi^g\rangle_{i,j,k}$ (tetramer ground state) is given by
\begin{equation}
   |\phi^g\rangle_{i,j,k}=\frac{1}{\sqrt{3}}
   \left(
   |T^+\rangle_{i,j}|t^-\rangle_k+|T^-\rangle_{i,j}|t^+\rangle_k
   -|T^0\rangle_{i,j}|t^0\rangle_k
   \right),
\label{eq:3} 
\end{equation}
where $\{|T^+\rangle_{i,j}, |T^0\rangle_{i,j}, |T^-\rangle_{i,j}\}$ represents the triplet states of the edge spins, i.e.,
\begin{equation}
 |T^{\alpha}\rangle_{i,j}=\begin{cases}
|\!\uparrow_{i}\uparrow_{j}\rangle & \mbox{ $(\alpha=+)$} \\
\left(|\!\uparrow_{i}\downarrow_{j}\rangle+|\!\downarrow_{i}\uparrow_{j}\rangle\right)/\sqrt{2}   & \mbox{ $(\alpha=0)$} \\
|\!\downarrow_{i}\downarrow_{j}\rangle           &  \mbox{ $(\alpha=-)$}
\end{cases},
\label{eq:4}
\end{equation}
and $\{|t^+\rangle_k, |t^0\rangle_k, |t^-\rangle_k\}$ represents the triplet states of a bond spin-pair, i.e.,
\begin{equation}
 |t^{\alpha}\rangle_{k}=\begin{cases}
|\!\uparrow_{k,a}\uparrow_{k,b}\rangle & \mbox{ $(\alpha=+)$} \\
\left(|\!\uparrow_{k,a}\downarrow_{k,b}\rangle+|\!\downarrow_{k,a}\uparrow_{k,b}\rangle\right)/\sqrt{2}   & \mbox{ $(\alpha=0)$} \\
|\!\downarrow_{k,a}\downarrow_{k,b}\rangle         &  \mbox{ $(\alpha=-)$}
\end{cases}.
\label{eq:5}
\end{equation}
Note that the bond spin-pair is in triplet states, but Eq. (\ref{eq:3}) is a nonmagnetic tetramer {\it{singlet}} state.

On the other hand, when the bond spin-pair is in a singlet state, the four interactions $J(=1)$ in the diamond unit vanish effectively. Therefore, the eigenfunctions of $h_{i,j}$ are simple product states, $|\sigma,\sigma'\rangle_{i,j}|s\rangle_k$, where $\sigma,\sigma'=\uparrow$ or $\downarrow$ and $|s\rangle_k$ represents the singlet state of the bond spin-pair\cite{Hirose2016}.

\subsection{Hamiltonian}

We consider the Hamiltonian on a diamond-like-decorated square lattice before introducing the further neighbor couplings. The Hamiltonian can be written by
\begin{equation}
   H_0=\sum_{\langle i,j \rangle}h_{i,j},
   \label{eq:6}
\end{equation}
where $\langle i,j \rangle$ represents a nearest-neighbor pair of the square lattice in Fig.~\ref{fig:2}(b).
When we regard the tetramer ground state $\phi^g$ as a dimer, the ground states of $H_0$ for $0.974<\lambda<2$, i.e., the MDTD states, are equivalent to the dimer-covering states of the square lattice\cite{Hirose2016,Hirose2017}.
In this study, we investigate the effects of further neighbor couplings $\Delta_{\rm{I}}$ and $\Delta_{\rm{II}}$, as shown in Fig.~\ref{fig:3}.
The coupling $\Delta_{\rm{I}}$ connects two adjacent diamond units in one plaquette and the coupling $\Delta_{\rm{II}}$ connects two diamond units facing each other in one plaquette.
We write the perturbation Hamiltonian as
\begin{equation}
   H'=\sum_{\langle k,k' \rangle}\Delta_{k,k'}(\mib{s}_{k,a}\cdot\mib{s}_{k',a}+\mib{s}_{k,b}\cdot\mib{s}_{k',b})
   \label{eq:7},
\end{equation}
\begin{equation}
   \Delta_{k,k'}=\begin{cases}
			\Delta_{\rm{I}}&\;\text{in the case where $\langle k,k' \rangle$ is a dashed line,}\\
			\Delta_{\rm{II}}&\;\text{in the case where $\langle k,k' \rangle$ is a double dashed line,}
			\end{cases}
\label{eq:8}																
\end{equation}
in Fig.~\ref{fig:3}.

\subsection{Matrix elements of a perturbation bond}

We define the perturbation operator between bond spin-pairs at sites $k$ and $k'$ as
\begin{equation}
   V_{k,k'}=\Delta_{k,k'}\left(\mib{s}_{k,a}\cdot\mib{s}_{k',a}+\mib{s}_{k,b}\cdot\mib{s}_{k'b}\right)
   \label{eq:9}.
\end{equation}
In the case where $\langle k,k'\rangle$ represents a $\Delta_{\rm{I}}$ bond and a $\Delta_{\rm{II}}$ bond, the states $|s\rangle_k|s\rangle_{k'}$ and $|t^{\alpha}\rangle_k|s\rangle_{k'}$ ($\alpha=-,\;0$, or $+$) can be operated on by the perturbation operator $V_{k,k'}$, and we have
\begin{equation}
   V_{k,k'}|s\rangle_k|s\rangle_{k'}=\frac{\Delta_{k,k'}}{2}\left(|t^0\rangle_k|t^0\rangle_{k'}-|t^+\rangle_k|t^-\rangle_{k'}-|t^-\rangle_k|t^+\rangle_{k'}\right)
   \label{eq:10},
\end{equation}
\begin{equation}
   V_{k,k'}|t^{\alpha}\rangle_k|s\rangle_{k'}=\frac{\Delta_{k,k'}}{2}|s\rangle_k|t^{\alpha}\rangle_{k'}.
\label{eq:11}
\end{equation}
On the other hand, in the case where $\langle k,k'\rangle$ represents only a $\Delta_{\rm{II}}$ bond, the states $|t^{\alpha}\rangle_k|t^{\alpha'}\rangle_{k'}$ ($\alpha, \alpha'=-,\;0$, or $+$) can be operated on by the perturbation operator $V_{k,k'}$, and we have
\begin{align}
V_{k,k'}|t^+\rangle_k|t^+\rangle_{k'}=&\frac{\Delta_{\rm{II}}}{2}|t^+\rangle_k|t^+\rangle_{k'}
\label{eq:12},\\
V_{k,k'}|t^-\rangle_k|t^-\rangle_{k'}=&\frac{\Delta_{\rm{II}}}{2}|t^-\rangle_k|t^-\rangle_{k'}
\label{eq:13},\\
V_{k,k'}|t^+\rangle_k|t^0\rangle_{k'}=&\frac{\Delta_{\rm{II}}}{2}|t^0\rangle_k|t^+\rangle_{k'}
\label{eq:14},\\
V_{k,k'}|t^-\rangle_k|t^0\rangle_{k'}=&\frac{\Delta_{\rm{II}}}{2}|t^0\rangle_k|t^-\rangle_{k'}
\label{eq:15},\\
V_{k,k'}|t^+\rangle_k|t^-\rangle_{k'}=&-\frac{\Delta_{\rm{II}}}{2}\left(|s\rangle_k|s\rangle_{k'}+|t^+\rangle_k|t^-\rangle_{k'}-|t^0\rangle_k|t^0\rangle_{k'}\right)
\label{eq:16},\\
V_{k,k'}|t^0\rangle_k|t^0\rangle_{k'}=&\frac{\Delta_{\rm{II}}}{2}\left(|s\rangle_k|s\rangle_{k'}+|t^+\rangle_k|t^-\rangle_{k'}+|t^-\rangle_k|t^+\rangle_{k'}\right)
\label{eq:17}.
\end{align}
Note that Eqs. (\ref{eq:12})--(\ref{eq:17}) are not related to a $\Delta_{\rm{I}}$ bond because a square-lattice site of dimer-covering states is prohibited from belonging to two or more dimers.

Note that Eqs. (\ref{eq:10}) and (\ref{eq:11}) do not yield the first-order term of the effective Hamiltonian because the right-hand sides of these equations do not contain diagonal terms. In our previous study, because only the couplings $\Delta_{\rm{I}}$ were introduced, we considered only Eqs. (\ref{eq:10}) and (\ref{eq:11}) and started with the second-order term. On the other hand, the right-hand sides of Eqs. (\ref{eq:12}), (\ref{eq:13}), and (\ref{eq:16}) contain diagonal terms and thus, one may think that these equations yield the first-order term of the effective Hamiltonian. However, in this case as well, it can easily be shown that the first-order term is zero, as follows. When we consider the tetramer singlet state $|\phi^g\rangle_{i,j,k}$, the expectation of the spin operator $s_{k,\nu}^{\xi}$ $(\xi=x,y,z$ ; $\nu=a,b)$ of the state $|\phi^g\rangle_{i,j,k}$ is obtained as $\langle\phi^g_{i,j,k}|s_{k,\nu}^{\xi}|\phi^g_{i,j,k}\rangle=0$. Therefore, we obtain $\langle\phi^g_{i,j,k};\phi^g_{i',j',k'}|V_{k,k'}|\phi^g_{i,j,k};\phi^g_{i',j',k'}\rangle=0$ and find that the first-order term of the effective Hamiltonian is not yielded.

\section{Second-Order Perturbation}\label{sec:3}

\subsection{Effective Hamiltonian}

We consider the second-order effective Hamiltonian.
Because possible second-order processes are created by using the perturbation bonds on a plaquette twice,
the second-order effective Hamiltonian can be written in the following form:
\begin{equation}
   H_{\rm{eff}}=-t \hat{T}+\epsilon_2\hat{D}_2+\epsilon_1\hat{D}_1+\epsilon_0\hat{D}_0,
\label{eq:18}   
\end{equation}
where $t$ represents the second-order pair-hopping amplitude and $\epsilon_2$, $\epsilon_1$, and $\epsilon_0$ represent the second-order perturbation energies when there are two, one, and zero dimers on a plaquette, respectively.
The operators $\hat{T}$, $\hat{D}_2$, $\hat{D}_1$, and $\hat{D}_0$ are defined by
\begin{equation}
\begin{minipage}[c]{.80\textwidth}
\begin{center}
\includegraphics[width=.63\linewidth]{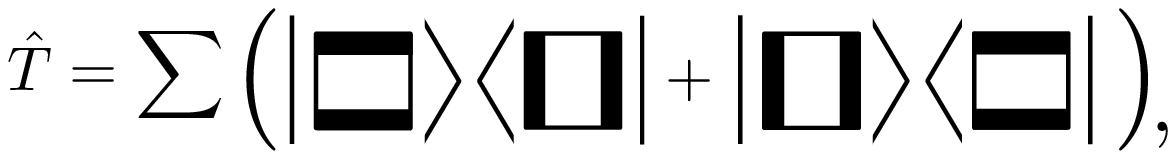}
\end{center}
\end{minipage}
\label{eq:19}
\end{equation}
\begin{equation}
\begin{minipage}[c]{.90\textwidth}
\begin{center}
\includegraphics[width=.58\linewidth]{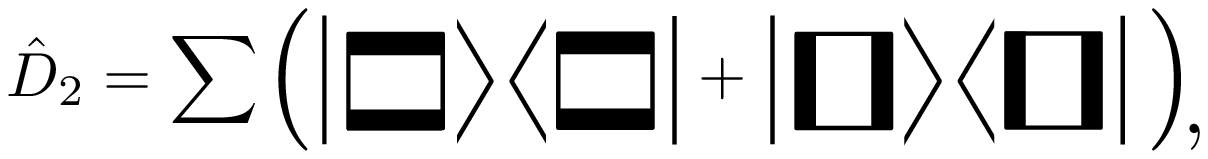}
\end{center}
\end{minipage}
\label{eq:20}
\end{equation}
\begin{equation}
\begin{minipage}[c]{.9\textwidth}
\begin{center}
\includegraphics[width=.95\linewidth]{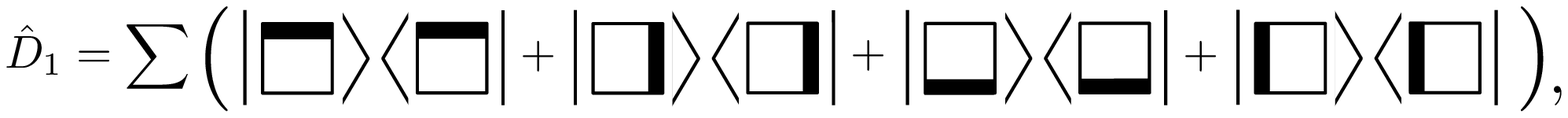}
\end{center}
\end{minipage}
\label{eq:21}
\end{equation}
\begin{equation}
\begin{minipage}[c]{.85\textwidth}
and
\begin{center}
\includegraphics[width=.35\linewidth]{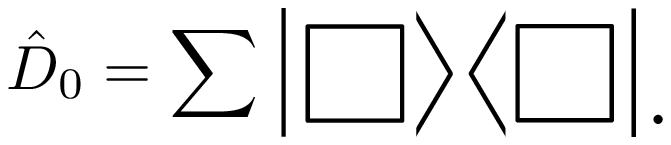}
\end{center}
\end{minipage}
\label{eq:22}
\end{equation}
We here use the conditions
\begin{equation}
   \hat{D}_2+\hat{D}_1+\hat{D}_0=N=\mbox{[total number of plaquettes]}
   \label{eq:23}
\end{equation}
and
\begin{equation}
   \frac{1}{2}\left(2\hat{D}_2+\hat{D}_1\right)=\frac{N}{2}=\mbox{[total number of dimers]}
   \label{eq:24}.
\end{equation}
Note that the coefficient of 1/2 on the left-hand side of Eq. (\ref{eq:24}) prevents double counting of the dimers.
Substituting Eqs. (\ref{eq:23}) and (\ref{eq:24}) into Eq. (\ref{eq:18}) and eliminating $\hat{D}_1$ and $\hat{D}_0$ from $H_{\rm{eff}}$, we obtain
\begin{equation}
   H_{\rm{eff}}= -t \hat{T}+(\epsilon_2-2\epsilon_1+\epsilon_0)\hat{D}_2+\epsilon_1N
   \label{eq:25}.
\end{equation}
Then, the coefficient of $\hat{D}_2$ on the right-hand side of Eq. (\ref{eq:25}) represents the dimer--dimer interaction
\begin{equation}
   v=\epsilon_2-2\epsilon_1+\epsilon_0.
\label{eq:26}   
\end{equation}
Equation (\ref{eq:26}) represents the repulsion ($v>0$) or attractive interaction ($v<0$) between dimers. Because the second-order perturbation energies $\epsilon_2$, $\epsilon_1$, and $\epsilon_0$ are negative, there is a large (small) energy gain of a plaquette with one dimer, i.e., $|\epsilon_1|\gg|\epsilon_0|, |\epsilon_2|$ ($|\epsilon_1|\ll|\epsilon_0|, |\epsilon_2|$), and the dimer--dimer interaction $v$ becomes repulsive (attractive). 
Then, we notice that the effective Hamiltonian $H_{\rm{eff}}$ can be written as the sum of $H_{\rm{QDM}}$ and the constant term $\epsilon_1N$,  which is the generation energy of a dimer.

Note that the above $t$, $v$, $\epsilon_2$, $\epsilon_1$, and $\epsilon_0$ are total second-order perturbation matrix elements when both couplings $\Delta_{\rm{I}}$ and $\Delta_{\rm{II}}$ are introduced. Then, we define the dimer--dimer interaction when we only introduce the couplings $\Delta_{\rm{I}}$ ($\Delta_{\rm{II}}$) as $v^{(\rm{I})}$ ($v^{(\rm{II})}$), and we can write
\begin{equation}
  v=v^{(\rm{I})}+v^{(\rm{II})}
\label{eq:27}.   
\end{equation}
Similarly, we define $\epsilon_2$, $\epsilon_1$, $\epsilon_0$, and $t$ as 
\begin{equation}
  \epsilon_2=\epsilon_2^{(\rm{I})}+\epsilon_2^{(\rm{II})}
\label{eq:28},   
\end{equation}
\begin{equation}
  \epsilon_1=\epsilon_1^{(\rm{I})}+\epsilon_1^{(\rm{II})}
\label{eq:29},   
\end{equation}
\begin{equation}
  \epsilon_0=\epsilon_0^{(\rm{I})}+\epsilon_0^{(\rm{II})}
\label{eq:30},   
\end{equation}
and
\begin{equation}
  t=t^{(\rm{I})}+t^{(\rm{II})}
\label{eq:31},
\end{equation}
because there is no cross term between the couplings $\Delta_{\rm{I}}$ and $\Delta_{\rm{II}}$ in the matrix elements $\epsilon_2$, $\epsilon_1$, $\epsilon_0$, and $t$.
Therefore, $v^{(\rm{I})}$ and $v^{(\rm{II})}$ can be written by
\begin{equation}
 v^{(\rm{I})}=\epsilon_2^{(\rm{I})}-2\epsilon_1^{(\rm{I})}+\epsilon_0^{(\rm{I})}
\label{eq:32}
\end{equation}
and
\begin{equation}
 v^{(\rm{II})}=\epsilon_2^{(\rm{II})}-2\epsilon_1^{(\rm{II})}+\epsilon_0^{(\rm{II})}
\label{eq:33}.
\end{equation}
Thus, the dimer--dimer interaction $v$ can be obtained by substituting Eqs. (\ref{eq:32}) and (\ref{eq:33}) into Eq. (\ref{eq:27}).

\subsection{Calculation results for dependence of $v/t$ on $\lambda$}

\begin{figure}[h]
\begin{center}
\includegraphics[width=.80\linewidth]{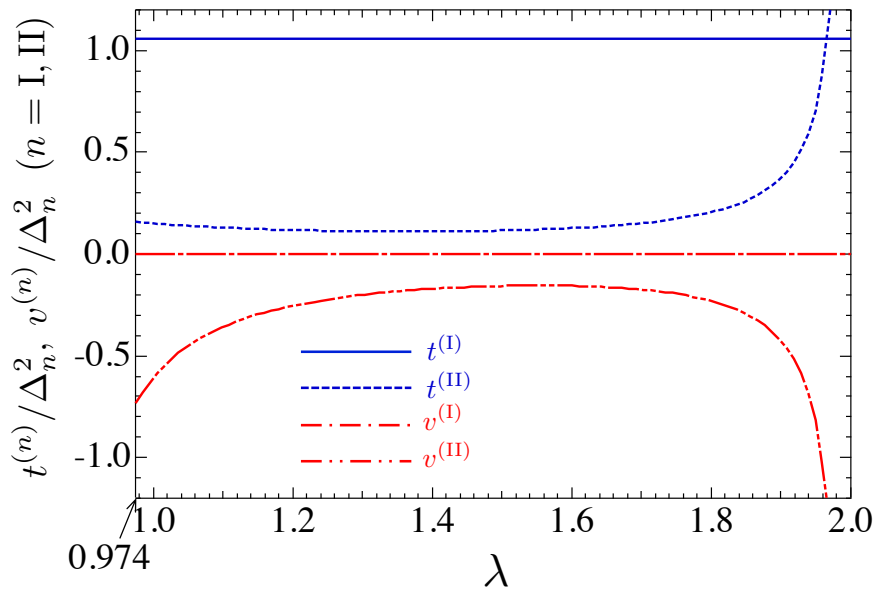}
\end{center}
\caption{(Color online) Calculation results for dependence of $v^{(\rm{I})}$, $t^{(\rm{I})}$, $v^{(\rm{II})}$, and $t^{(\rm{II})}$ on $\lambda$.}
\label{fig:4}
\end{figure}

In Fig.~\ref{fig:4}, we show the numerical calculation results for the dependence of $v^{(\rm{I})}$, $t^{(\rm{I})}$, $v^{(\rm{II})}$, and $t^{(\rm{II})}$ on $\lambda$. 
We take $0.974<\lambda<2$ for the horizontal axis, where the MDTD states are stabilized and the square lattice dimer-covering states are constructed.
The dimer--dimer interaction $v^{(\rm{I})}$ becomes zero in the whole region of $\lambda$, which is based on the fact that the contributions of the perturbation process cancel each other out\cite{Hirose2016} (see Appendix A.1).
On the other hand, the dimer--dimer interaction $v^{(\rm{II})}$ becomes $v^{(\rm{II})}\neq0$ because the contributions of the perturbation process do not cancel each other out, and we obtain $v^{(\rm{II})}<0$, which is the attractive interaction between dimers,  in the whole region of $\lambda$. 
Furthermore, the results of $v^{(\rm{I})}=0$ and $v^{(\rm{II})}<0$ indicate that instead of the coupling $\Delta_{\rm{I}}$, the coupling $\Delta_{\rm{II}}$ produces an attractive interaction. 
The large attractive interaction $v^{(\rm{II})}$ in the neighborhood of $\lambda=0.974$ ($\lambda=2$) originates from the fact that the energy gain of a plaquette with no (two) dimers is larger than the others, i.e., $|\epsilon_0|\gg|\epsilon_1|, |\epsilon_2|$ ($|\epsilon_2|\gg|\epsilon_0|, |\epsilon_1| $). In the intermediate region of $\lambda$, the energy gains of a plaquette with two, one, and no dimers are comparable, which produce a weak attractive interaction.  These details are described in Appendix A.2.
Then, focusing on the hopping parameter, $t^{(\rm{I})}$ has no dependence on $\lambda$ and becomes $t^{(\rm{I})}/\Delta_{\rm{I}}^2=1.06$ \cite{Hirose2016}; on the other hand, $t^{(\rm{II})}$ depends on $\lambda$. 
The dependence on $\lambda$ originates from the difference between the number of dimers in the intermediate state and that in the initial (final) state in the perturbation process. If the numbers of dimers are the same in both the initial (final) and intermediate states, there is no $\lambda$ dependence. On the other hand, if the numbers of dimers are different in these states, there is $\lambda$ dependence. We describe the details of these dependences in Appendix A.3 and A.4.
Furthermore, $v^{(\rm{II})}$ and $t^{(\rm{II})}$ diverge to $-\infty$ and $+\infty$, respectively, for $\lambda\to2$.
Because the ground state becomes the dimer--monomer state for $\lambda\ge 2$ in the original spin Hamiltonian\cite{Hirose2017}, the point $\lambda=2$ is a phase transition point and the energy denominator becomes zero.

\begin{figure}[h]
\begin{center}
\includegraphics[width=.80\linewidth]{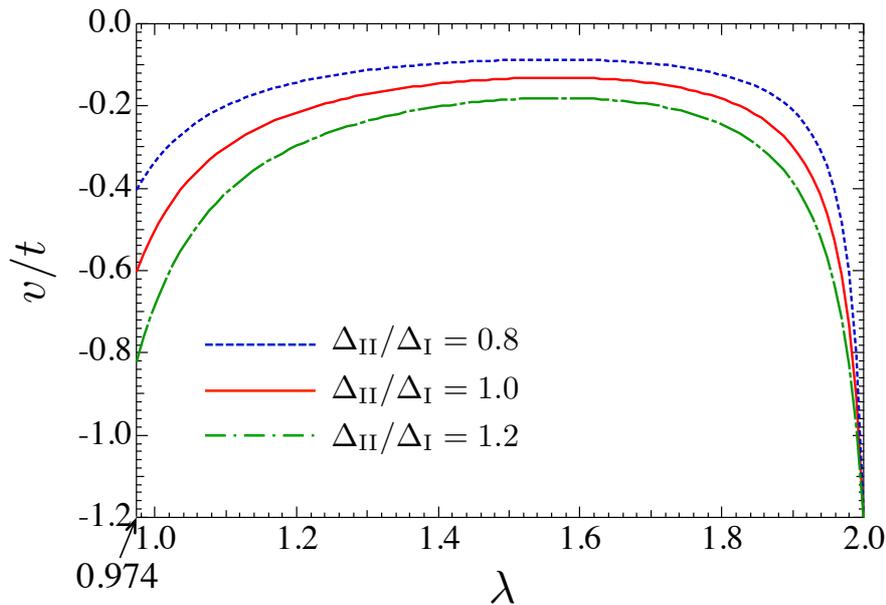}
\end{center}
\caption{(Color online) Calculation results for the dependence of $v/t$ on $\lambda$ in the cases of $\Delta_{\rm{II}}/\Delta_{\rm{I}}=0.8, 1.0,$ and $1.2$.}
\label{fig:5}
\end{figure}

By substituting $v^{(\rm{I})}(=0)$, $t^{(\rm{I})}$, $v^{(\rm{II})}(<0)$, and $t^{(\rm{II})}$ into Eqs. (\ref{eq:27}) and (\ref{eq:31}), we can obtain $v/t=\frac{v^{(\rm{II})}}{t^{(\rm{I})}+t^{(\rm{II})}}=\frac{a\Delta_{\rm{II}}^2}{(b\Delta_{\rm{I}})^2+(c\Delta_{\rm{II}})^2}(\le0)$, where $a, b,$ and $c$ are constant and $a$ has a negative value, which originates from $v^{(\rm{II})}<0$. Note that, in the case of $\Delta_{\rm{II}}=0$, we obtain $v/t=0$. In Fig.~\ref{fig:5}, we show the calculation results for the dependence of $v/t$ on $\lambda$ in the cases of $\Delta_{\rm{II}}/\Delta_{\rm{I}}=0.8, 1.0,$ and $1.2$.  
From Fig.~\ref{fig:5}, we can see that the result of $-1.2\le v/t\le0$ ($v\le0$, $t>0$) is obtained and that $|v/t|$ increases as $\Delta_{\rm{II}}/\Delta_{\rm{I}}$ becomes large, except in the neighborhood of $\lambda=2$. 
On the other hand, $|v/t|$ approaches zero as $\Delta_{\rm{II}}/\Delta_{\rm{I}}$ becomes small.
In all cases of $\Delta_{\rm{II}}/\Delta_{\rm{I}}=0.8, 1.0,$ and $1.2$, $v/t$ converges to $v/t=-1.2$ for $\lambda\to2$. 
On the other hand, $\lambda=0.974$ is the phase transition point between the MDTD and ferrimagnetic ground states in the original spin Hamiltonian, and $v/t$ smoothly decreases towards $\lambda=0.974$, in contrast to the sharp decrease at $\lambda=2$.
The result of $-1.2\le v/t\le0$ ($v\le0$, $t>0$) shows that the pair hopping of dimers occurs and that there is an attractive interaction between dimers. 
Because recent studies have shown that the columnar phase is stabilized for the attractive square-lattice QDM\cite{square_qdm_2006, square_qdm_2008, Moessner2008, book, square_qdm_2014, square_qdm_2016}, we suggest that our obtained result corresponds to the columnar phase.


\section{Summary}\label{sec:4}

We calculated the second-order effective Hamiltonian for spin-1/2 Heisenberg antiferromagnets on a diamond-like-decorated square lattice and derived the square-lattice QDM by introducing two kinds of further neighbor couplings $\Delta_{\rm{I}}$ and $\Delta_{\rm{II}}$.
In the case of introducing only the coupling $\Delta_{\rm{I}}$, our second-order effective Hamiltonian becomes a square-lattice QDM with a finite pair-hopping amplitude ($t>0$) and no dimer--dimer interaction ($v=0$), which was discussed in our previous paper\cite{Hirose2016}. 
On the other hand, when we introduce the coupling $\Delta_{\rm{II}}$ in addition to $\Delta_{\rm{I}}$, a negative dimer--dimer interaction ($v\le0$), i.e., an  attractive interaction between dimers, is generated. Therefore, we found that the coupling $\Delta_{\rm{II}}$ causes an attractive interaction between dimers.
As a result, in the case of introducing both couplings $\Delta_{\rm{I}}$ and $\Delta_{\rm{II}}$, we obtained the square-lattice QDM with $-1.2\le v/t\le0$ ($v\le0$, $t>0$). 
Our QDM can be realized experimentally in a quantum simulator using optical lattices\cite{Bloch2008} because our model contains only quadratic Heisenberg-type exchange terms. 
Furthermore, if our diamond-like-decorated square lattice is realized using optical lattices, we can perform a Raman scattering experiment to identify whether the square-lattice QDM is realized as the low-energy effective Hamiltonian because the singlet sector can be observed in a Raman scattering experiment\cite{raman1}. In order to obtain calculation data for comparison with the experimental results, we are also planning to perform calculations of the Raman spectrum\cite{raman2} on both the diamond-like-decorated square-lattice Heisenberg antiferromagnet and the square-lattice QDM by the numerical diagonalization of finite-size systems. Then, by reproducing the calculation situations with optical lattice systems and by comparing the Raman spectra, we can confirm the realization of the QDM in the optical lattice. 
We obtained the square-lattice QDM with $-1.2\le v/t\le0$, but it remains unclear which phase, the columnar phase, the plaquette phase, or the mixed columnar-plaquette phase, is stabilized in this region of weak dimer-dimer interaction\cite{square_qdm_1996, square_qdm_2006, square_qdm_2008, Moessner2008, book, square_qdm_2014, square_qdm_2016}. If we refer to the latest studies of the square-lattice QDM by Banerjee et al\cite{square_qdm_2014, square_qdm_2016}., our obtained result corresponds to the columnar phase. Therefore, we are very interested in examining which of the above three phases is stabilized in the QDM in the optical lattice. Thus, we are also planning to theoretically investigate the qualitative differences that will appear in the Raman spectrum in the above three phases. In a Raman scattering experiment, there is a degree of freedom of the electric field in the incident direction and that in the reflection direction; thus,  we are interested in whether there is a relationship between the directions of the electric field and the dimer arrangement in each phase.

Our results were obtained under the condition that the direction of the bond spin-pairs is orthogonal to the plane formed by the edge spins, as shown in Fig. {\ref{fig:2}}.
On the other hand, when we choose the direction of the bond spin-pairs to be parallel to the edge-spin plane, as shown in Fig. {\ref{fig:1}}, the matrix elements of a perturbation bond become more complicated because the further neighbor couplings are not symmetric with respect to $\mib{s}_{k,a}\leftrightarrow\mib{s}_{k,b}$. Therefore, the calculation results for the dependence of $v/t$ on $\lambda$ could be different from the results shown in Fig. {\ref{fig:5}}. However, we also show that in this case, the contributions of the perturbation process cancel each other out when we introduce only the coupling $\Delta_{\rm{I}}$, and that the coupling $\Delta_{\rm{II}}$ yields the dimer--dimer interaction.
Examining whether this dimer--dimer interaction becomes an attractive interaction, as shown in this study, or a repulsive interaction is a future issue.

\section*{Appendix: \;Calculations of the second-order perturbation matrix elements \\
\;\;\;\;\;\;\;\;\;\;\;\;\;\;\;\;\;\;\;$v^{(\rm{I})}$, $v^{(\rm{II})}$, $t^{(\rm{I})}$, and $t^{(\rm{II})}$.}

\setcounter{equation}{0} 
\renewcommand{\theequation}{A.\arabic{equation}}
\setcounter{figure}{0} 
\renewcommand{\thefigure}{A.\arabic{figure}}
Here, we describe the details of the calculations of the second-order perturbation matrix elements $v^{(\rm{I})}$, $v^{(\rm{II})}$, $t^{(\rm{I})}$, and $t^{(\rm{II})}$.

\subsection*{A.1\; Calculation process for the dimer--dimer interaction $v^{(\rm{I})}$}

In our previous study, we obtained
\begin{align}
\epsilon_1^{(\rm{I})}=\frac{\epsilon_0^{(\rm{I})}+\epsilon_2^{(\rm{I})}}{2}
\label{eq:34},
\end{align}
which is based on the existence of equivalent clusters in the intermediate states of the perturbation process [See Sect. 3.2 of Ref. 27]. Substituting Eq. (\ref{eq:34}) into Eq. (\ref{eq:32}), we obtain 
\begin{align}
v^{(\rm{I})}=0.
\label{eq:35}
\end{align}

\subsection*{A.2\; Calculation process for the dimer--dimer interaction $v^{(\rm{II})}$}

\begin{figure}[!htb]
\begin{center}
\includegraphics[width=.95\linewidth]{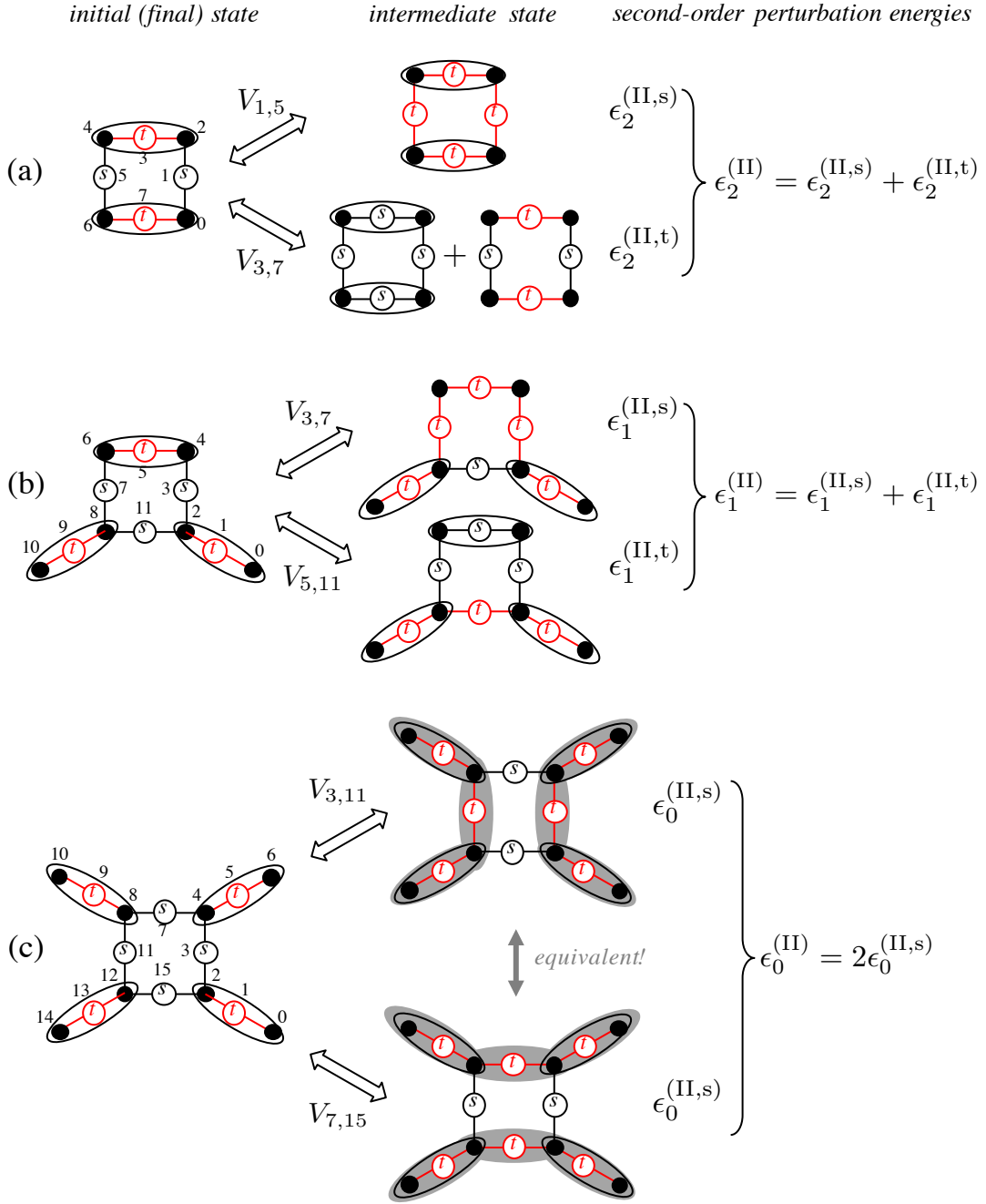}
\end{center}
\caption{(Color online) Second-order perturbation processes and energies when we use the coupling $\Delta_{\rm{II}}$ in the case where the initial state has (a) two dimers, (b) one dimer, and (c) no dimer on the plaquette.
The red bonds indicate that the bond spin-pair is in triplet states.
A red bond with an oval represents the tetramer ground state $\phi^g$.
A black bond with an oval represents the state that is obtained by replacing $|t\rangle$ in $\phi^g$ with $|s\rangle$.}
\label{fig:7}
\end{figure}

In Fig.~\ref{fig:7}, we show the possible second-order perturbation processes when we use the coupling $\Delta_{\rm{II}}$. 
In Fig.~\ref{fig:7}(a), the initial state has two dimers on the plaquette. There are two kinds of processes, which are produced by $V_{1,5}$ and $V_{3,7}$. When we use the operator $V_{1,5}$, the singlet states at sites $1$ and $5$ turn into triplet states in the intermediate state. On the other hand, when we use the operator $V_{3,7}$, a cluster where sites 3 and 7 are in triplet states and a cluster where they are in singlet states are generated in the intermediate state. Defining the second-order perturbation energy when we use the operator $V_{1,5}$ ($V_{3,7}$) as $\epsilon_2^{(\rm{II},\rm{s})}$ ($\epsilon_2^{(\rm{II},\rm{t})}$), we can write 
\begin{align}
\epsilon_2^{(\rm{II})}=\epsilon_2^{(\rm{II},\rm{s})}+\epsilon_2^{(\rm{II},\rm{t})}
\label{eq:36}.
\end{align}
In Fig.~\ref{fig:7}(b), the initial state has one dimer on the plaquette. There are also two kinds of processes, which are produced by $V_{3,7}$ and $V_{5,11}$. When we use the operator $V_{3,7}$, the singlet states at sites $3$ and $7$ become triplet states, and when we use the operator $V_{5,11}$, 
the triplet and singlet states at sites $5$ and $11$, respectively, are replaced by each other. Defining the second-order perturbation energy when we use the operator $V_{3,7}$ ($V_{5,11}$) as $\epsilon_1^{(\rm{II},\rm{s})}$ ($\epsilon_1^{(\rm{II},\rm{t})}$), we can write 
\begin{align}
\epsilon_1^{(\rm{II})}=\epsilon_1^{(\rm{II},\rm{s})}+\epsilon_1^{(\rm{II},\rm{t})}
\label{eq:37}.
\end{align}
In Fig.~\ref{fig:7}(c), the initial state has no dimer on the plaquette. In this case, when $V_{3,11}$ and $V_{7,15}$ operate separately, the clusters formed in the intermediate state become equivalent to each other. Therefore, we have the same contributions from the process produced by these operators. Thus, defining the second-order perturbation energy when we use the operator $V_{3,11}$ or $V_{7,15}$ as $\epsilon_0^{(\rm{II},\rm{s})}$, we can write 
\begin{align}
\epsilon_0^{(\rm{II})}=2\epsilon_0^{(\rm{II},\rm{s})}
\label{eq:38}.
\end{align}
Note that equivalent clusters in the intermediate states exist in Fig.~\ref{fig:7}(c) but do not exist between Figs.~\ref{fig:7}(a), \ref{fig:7}(b), and \ref{fig:7}(c).
Therefore, because there is no relationship between $\epsilon_2^{(\rm{II})}$, $\epsilon_1^{(\rm{II})}$, and $\epsilon_0^{(\rm{II})}$ such as Eq. (\ref{eq:34}), we have $v^{(\rm{II})}\neq0$. 

\begin{figure}[h]
\begin{center}
\includegraphics[width=.80\linewidth]{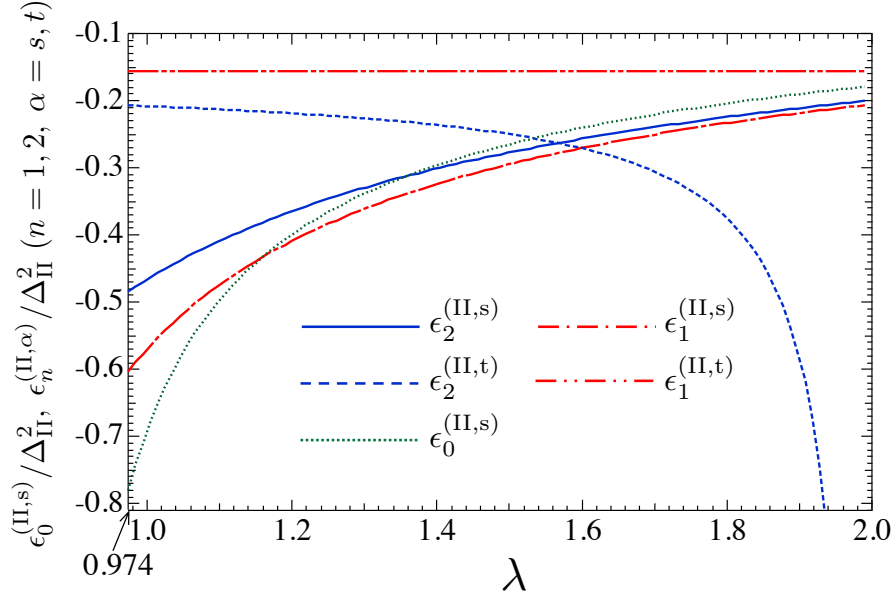}
\end{center}
\caption{(Color online) Calculation results for the dependence of $\epsilon_2^{\rm{(II,s)}}$, $\epsilon_2^{\rm{(II,t)}}$, $\epsilon_1^{\rm{(II,s)}}$, $\epsilon_1^{\rm{(II,t)}}$, and $\epsilon_0^{\rm{(II,s)}}$ on $\lambda$.}
\label{fig:8}
\end{figure}

In Fig.~\ref{fig:8}, we show the numerical calculation results for the dependence of $\epsilon_2^{\rm{(II,s)}}$, $\epsilon_2^{\rm{(II,t)}}$, $\epsilon_1^{\rm{(II,s)}}$, $\epsilon_1^{\rm{(II,t)}}$, and $\epsilon_0^{\rm{(II,s)}}$ on $\lambda$. These results show that the energy gain of a plaquette with no dimer, $\epsilon_0^{\rm{(II,s)}}$, is larger than the others in the neighborhood of $\lambda=0.974$. Therefore, in Fig.~\ref{fig:4}, $v^{(\rm{II})}$ becomes a large attractive interaction at $\lambda=0.974$. On the other hand, in the neighborhood of $\lambda=2$, the energy gain of a plaquette with two dimers, $\epsilon_2^{\rm{(II,t)}}$, is much larger than the others and $\epsilon_2^{\rm{(II,t)}}$ diverges to $-\infty$ at $\lambda=2$, which produces the divergence of $v^{\rm{(II)}}\to-\infty$ at $\lambda=2$, as shown in Fig.~\ref{fig:4}. The reason for the divergence of $v^{\rm{(II)}}$ at $\lambda=2$ is discussed later. Furthermore, in the intermediate region of $\lambda$, we find that all energy gains are comparable; therefore, $v^{(\rm{II})}$ becomes a small attractive interaction, as shown in Fig.~\ref{fig:4}.  The value of $v^{(\rm{II})}$ can be obtained by calculating the right-hand sides of Eqs. (\ref{eq:36})--(\ref{eq:38}) and substituting them into Eq. (\ref{eq:33}).

\subsubsection*{A.2.1\; Calculation of the energy $\epsilon_2^{(\rm{II})}$}

First, we consider the upper process $\epsilon_2^{(\rm{II},\rm{s})}$ in Fig.~\ref{fig:7}(a). When $V_{1,5}$ operates on the initial state $|s_{1};\phi^g_{2,4,3};s_{5};\phi^g_{0,6,7}\rangle$, we have 
\begin{equation}
   V_{1,5}|s_{1};\phi^g_{2,4,3};s_{5};\phi^g_{0,6,7}\rangle=\frac{\Delta_{\rm{II}}}{2}\left(|t^0\rangle_1|t^0\rangle_{5}-|t^+\rangle_1|t^-\rangle_{5}-|t^-\rangle_1|t^+\rangle_{5}\right)|\phi^g_{2,4,3};\phi^g_{0,6,7}\rangle
   \label{eq:39},
\end{equation}
where we use Eq. (\ref{eq:10}). If the bond spin-pairs at sites $1$ and $5$ are in triplet states, then a connected cluster $(0,1,2,3,4,5,6,7)$ appears in the intermediate state. 
For the unperturbed Hamiltonian, $h_{0,2}+h_{2,4}+h_{4,6}+h_{6,0}$ for this cluster, and we write the eigenvalues and eigenstates as $E_{0-7}^n+4\lambda$ and $|\Psi^n\rangle_{0-7}$, respectively.
Therefore, the matrix elements are written by 
\begin{align}
\langle&\Psi^n_{0-7}|V_{1,5}|s_{1};\phi^g_{2,4,3};s_{5};\phi^g_{0,6,7}\rangle \notag\\
=&\frac{\Delta_{\rm{II}}}{2}\left(\langle\Psi^n_{0-7}|t_1^0;t_5^0;\phi^g_{2,4,3};\phi^g_{0,6,7}\rangle-\langle\Psi^n_{0-7}|t_1^+;t_5^-;\phi^g_{2,4,3};\phi^g_{0,6,7}\rangle-\langle\Psi^n_{0-7}|t_1^-;t_5^+;\phi^g_{2,4,3};\phi^g_{0,6,7}\rangle\right)
 \label{eq:40}.
\end{align}
The energy denominator of the intermediate state is given by
\begin{align}
E_{0-7}^n+4\lambda-2(\lambda-2)=E_{0-7}^n+2\lambda+4
 \label{eq:41}.
\end{align}
Thus, from Eqs. (\ref{eq:40}) and (\ref{eq:41}), we obtain
\begin{align}
\epsilon_2^{(\rm{II},\rm{s})}=-\frac{\Delta_{\rm{II}}^2}{4}\sum_{n}\frac{{\left(\langle\Psi^n_{0-7}|t_1^0;t_5^0;\phi^g_{2,4,3};\phi^g_{0,6,7}\rangle-\langle\Psi^n_{0-7}|t_1^+;t_5^-;\phi^g_{2,4,3};\phi^g_{0,6,7}\rangle-\langle\Psi^n_{0-7}|t_1^-;t_5^+;\phi^g_{2,4,3};\phi^g_{0,6,7}\rangle\right)}^2}{E_{0-7}^n+2\lambda+4}
\label{eq:42}.
\end{align}

Next, we consider the lower process $\epsilon_2^{(\rm{II},\rm{t})}$ in Fig.~\ref{fig:7}(a). When $V_{3,7}$ operates on the initial state $|\phi^g_{2,4,3};\phi^g_{0,6,7}\rangle$, we have 
\begin{align}
V&_{3,7}|\phi^g_{2,4,3};\phi^g_{0,6,7}\rangle \notag \\
=&-\frac{\Delta_{\rm{II}}}{6}\left(|T^+\rangle_{2,4}|s\rangle_{3}|T^-\rangle_{0,6}|s\rangle_{7}+|T^-\rangle_{2,4}|s\rangle_{3}|T^+\rangle_{0,6}|s\rangle_{7}-|T^0\rangle_{2,4}|s\rangle_{3}|T^0\rangle_{0,6}|s\rangle_{7}\right)\notag \\
&+\frac{\Delta_{\rm{II}}}{6}\Bigl[|T^+\rangle_{2,4}\left(|t^-\rangle_{3}|T^+\rangle_{0,6}|t^-\rangle_{7}-|t^-\rangle_{3}|T^-\rangle_{0,6}|t^+\rangle_{7}+|t^0\rangle_{3}|T^-\rangle_{0,6}|t^0\rangle_{7}-|t^0\rangle_{3}|T^0\rangle_{0,6}|t^-\rangle_{7}\right) \notag \\
&\;\;\;\;+|T^-\rangle_{2,4}\left(-|t^+\rangle_{3}|T^+\rangle_{0,6}|t^-\rangle_{7}+|t^0\rangle_{3}|T^+\rangle_{0,6}|t^0\rangle_{7}+|t^+\rangle_{3}|T^-\rangle_{0,6}|t^+\rangle_{7}-|t^0\rangle_{3}|T^0\rangle_{0,6}|t^+\rangle_{7}\right) \notag \\
&\;\;\;\;-|T^0\rangle_{2,4}\left(|t^-\rangle_{3}|T^+\rangle_{0,6}|t^0\rangle_{7}+|t^+\rangle_{3}|T^-\rangle_{0,6}|t^0\rangle_{7}-|t^+\rangle_{3}|T^0\rangle_{0,6}|t^-\rangle_{7}-|t^-\rangle_{3}|T^0\rangle_{0,6}|t^+\rangle_{7}\right)
\Bigr].
\label{eq:43}
\end{align}
In the first term on the right-hand side of Eq. (\ref{eq:43}), the bond spin-pairs at sites $3$ and $7$ are in the singlet states; on the other hand, in the second term, they are in the triplet states.
First, we consider the case where they are in singlet states in the intermediate state. For the unperturbed Hamiltonian $h_{2,4}+h_{0,6}$, defining the eigenstate as $|S^n\rangle_{0,2,3,4,6,7}\equiv|s_3;s_7;\sigma_{0};\sigma_{2};\sigma_{4};\sigma_{6}\rangle$ with $\sigma_{\xi}=\uparrow, \downarrow$ $(\xi=0,2,4,6)$, the matrix elements are written by
\begin{align}
\langle S&_{0,2,3,4,6,7}^n|V_{3,7}|\phi^g_{2,4,3};\phi^g_{0,6,7}\rangle \notag \\
=&\langle s_3;s_7;\sigma_{0};\sigma_{2};\sigma_{4};\sigma_{6}|V_{3,7}|\phi^g_{2,4,3};\phi^g_{0,6,7}\rangle \notag \\
=&-\frac{\Delta_{\rm{II}}}{6}\left(\delta_{\sigma_0,\downarrow}\delta_{\sigma_2,\uparrow}\delta_{\sigma_4,\uparrow}\delta_{\sigma_6,\downarrow}+\delta_{\sigma_0,\uparrow}\delta_{\sigma_2,\downarrow}\delta_{\sigma_4,\downarrow}\delta_{\sigma_6,\uparrow}\right) \notag \\
&\;+\frac{\Delta_{\rm{II}}}{12}\left(\delta_{\sigma_0,\uparrow}\delta_{\sigma_2,\uparrow}\delta_{\sigma_4,\downarrow}\delta_{\sigma_6,\downarrow}
+\delta_{\sigma_0,\downarrow}\delta_{\sigma_2,\uparrow}\delta_{\sigma_4,\downarrow}\delta_{\sigma_6,\uparrow}
+\delta_{\sigma_0,\uparrow}\delta_{\sigma_2,\downarrow}\delta_{\sigma_4,\uparrow}\delta_{\sigma_6,\downarrow}
+\delta_{\sigma_0,\downarrow}\delta_{\sigma_2,\downarrow}\delta_{\sigma_4,\uparrow}\delta_{\sigma_6,\uparrow}\right)
\label{eq:44}.
\end{align}
Because the eigenvalues for $|S^n\rangle_{0,2,3,4,6,7}$ are zero, the energy denominator of the intermediate state is given by
\begin{align}
0-2(\lambda-2)=4-2\lambda
 \label{eq:45}.
\end{align}
Therefore, defining the perturbation energy when the singlet states at sites $3$ and $7$ are formed in the intermediate state as $\epsilon_2^{(\rm{II},\rm{t},\alpha)}$, we obtain
\begin{align}
\epsilon_2^{(\rm{II},\rm{t},\alpha)}=&-\frac{\Delta_{\rm{II}}^2}{4-2\lambda}\sum_n{|\langle S_{0,2,3,4,6,7}^n|V_{3,7}|\phi^g_{2,4,3};\phi^g_{0,6,7}\rangle|}^2 \notag \\
=&-\frac{\Delta_{\rm{II}}^2}{4-2\lambda}\sum_{\sigma_{0}=\uparrow,\downarrow}\sum_{\sigma_{2}=\uparrow,\downarrow}\sum_{\sigma_{4}=\uparrow,\downarrow}\sum_{\sigma_{6}=\uparrow,\downarrow}{|\langle s_3;s_7;\sigma_{0};\sigma_{2};\sigma_{4};\sigma_{6}|V_{3,7}|\phi^g_{2,4,3};\phi^g_{0,6,7}\rangle|}^2, \notag \\
=&\frac{\Delta_{\rm{II}}^2}{24(\lambda-2)}
 \label{eq:46},
\end{align}
where we find that the denominator of the right-hand side of Eq. (\ref{eq:46}) produces the divergence of $v^{(\rm{II})}\to-\infty$ at $\lambda=2$.

Next, we consider the case where the bond spin-pairs at sites 3 and 7 are in triplet states in the intermediate state.
For the unperturbed Hamiltonian $h_{2,4}+h_{0,6}$, we define the eigenstate as $|\phi_{2,4,3};\phi_{0,6,7}\rangle$, where $|\phi\rangle_{i,j,k}$ with $(i,j,k)=(2,4,3)$ and $(0,6,7)$ is the eigenvector for  $h_{i,j}$.
Below, we write the eigenvectors for each eigenvalue $e_{\lambda}$ for $h_{i,j}$.
For $e_{\lambda}=\lambda+1$, we obtain five degenerated states,
\begin{equation}
 |\phi\rangle_{i,j,k}=\begin{cases}
\frac{1}{\sqrt{6}}\left(|T^+\rangle_{i,j}|t^-\rangle_{k}+|T^-\rangle_{i,j}|t^+\rangle_{k}+2|T^0\rangle_{i,j}|t^0\rangle_{k}\right), \\
\frac{1}{\sqrt{2}} \left(|T^+\rangle_{i,j}|t^0\rangle_{k}+|T^0\rangle_{i,j}|t^+\rangle_{k}\right),\\
\frac{1}{\sqrt{2}} \left(|T^-\rangle_{i,j}|t^0\rangle_{k}+|T^0\rangle_{i,j}|t^-\rangle_{k}\right), \\
|T^+\rangle_{i,j}|t^+\rangle_{k},\\
|T^-\rangle_{i,j}|t^-\rangle_{k};
\end{cases}
\label{eq:47}
\end{equation}
for $e_{\lambda}=\lambda-1$, we find triple degenerated states,
\begin{equation}
 |\phi\rangle_{i,j,k}=\begin{cases}
\frac{1}{\sqrt{2}}\left(|T^+\rangle_{i,j}|t^-\rangle_{k}-|T^-\rangle_{i,j}|t^+\rangle_{k}\right), \\
\frac{1}{\sqrt{2}} \left(|T^+\rangle_{i,j}|t^0\rangle_{k}-|T^0\rangle_{i,j}|t^+\rangle_{k}\right),\\
\frac{1}{\sqrt{2}} \left(|T^-\rangle_{i,j}|t^0\rangle_{k}-|T^0\rangle_{i,j}|t^-\rangle_{k}\right);
\end{cases}
\label{eq:48}
\end{equation}
and for $e_{\lambda}=\lambda-2$, which is the ground-state energy, we find the tetramer singlet state in Eq. (\ref{eq:3}).
Therefore, because $|\phi\rangle_{i,j,k}$ has the nine states in Eqs. (\ref{eq:47}), (\ref{eq:48}), and (\ref{eq:3}), the intermediate state $|\phi_{2,4,3};\phi_{0,6,7}\rangle$ has $80$ states, where we subtract $1$ from $9^2$ because we exclude the state $|\phi_{2,4,3};\phi_{0,6,7}\rangle=|\phi^g_{2,4,3};\phi^g_{0,6,7}\rangle$.
Calculating the matrix elements for the $80$ states and defining the perturbation energy when the triplet states at sites $3$ and $7$ are formed in the intermediate state as $\epsilon_2^{(\rm{II},\rm{t},\beta)}$, we obtain
\begin{align}
\epsilon_2^{(\rm{II},\rm{t},\beta)}=-\frac{\Delta_{\rm{II}}^2}{6}
\label{eq:49}.
\end{align}
From Eqs. (\ref{eq:46}) and (\ref{eq:49}), $\epsilon_2^{(\rm{II},\rm{t})}$ is obtained by
\begin{align}
\epsilon_2^{(\rm{II},\rm{t})}=&\epsilon_2^{(\rm{II},\rm{t},\alpha)}+\epsilon_2^{(\rm{II},\rm{t},\beta)} \notag \\
						=&\left\{\frac{1}{24(\lambda-2)}-\frac{1}{6}\right\}\Delta_{\rm{II}}^2
\label{eq:50}.
\end{align}
Therefore, substituting Eqs. (\ref{eq:42}) and (\ref{eq:50}) into Eq. (\ref{eq:36}), we obtain $\epsilon_2^{(\rm{II})}$.

\subsubsection*{A.2.2 \;Calculation of the energy $\epsilon_1^{(\rm{II})}$}

First, we consider the upper process $\epsilon_1^{(\rm{II},\rm{s})}$ in Fig.~\ref{fig:7}(b). When $V_{3,7}$ operates on the initial state $|\phi^g_{0,2,1};s_{3};\phi^g_{4,6,5};s_{7};\phi^g_{8,10,9}\rangle$, we have 
\begin{align}
   V_{3,7}|\phi^g_{0,2,1};s_{3}&;\phi^g_{4,6,5};s_{7};\phi^g_{8,10,9}\rangle=&\frac{\Delta_{\rm{II}}}{2}\left(|t^0\rangle_3|t^0\rangle_{7}-|t^+\rangle_3|t^-\rangle_{7}-|t^-\rangle_3|t^+\rangle_{7}\right)|\phi^g_{0,2,1};\phi^g_{4,6,5};\phi^g_{8,10,9}\rangle
   \label{eq:51}.
\end{align}
If the bond spin-pairs at sites $3$ and $7$ are in triplet states, then a connected cluster $(0,1,2,3,4,5,6,7,8,9,10)$ appears in the intermediate state. 
For the unperturbed Hamiltonian, $h_{0,2}+h_{2,4}+h_{4,6}+h_{6,8}+h_{8,10}$ for this cluster, and we write the eigenvalues and eigenstates as $E_{0-10}^n+5\lambda$ and $|\Psi^n\rangle_{0-10}$, respectively.
Therefore, the matrix elements are written by 
\begin{align}
\langle\Psi^n_{0-10}|V_{3,7}|\phi^g_{0,2,1}&;s_{3};\phi^g_{4,6,5};s_{7};\phi^g_{8,10,9}\rangle \notag\\
=&\frac{\Delta_{\rm{II}}}{2}\bigl(\langle\Psi^n_{0-10}|t_3^0;t_7^0;\phi^g_{0,2,1};\phi^g_{4,6,5};\phi^g_{8,10,9}\rangle-\langle\Psi^n_{0-10}|t_3^+;t_7^-;\phi^g_{0,2,1};\phi^g_{4,6,5};\phi^g_{8,10,9}\rangle \notag \\
&\;\;\;-\langle\Psi^n_{0-10}|t_3^-;t_7^+;\phi^g_{0,2,1};\phi^g_{4,6,5};\phi^g_{8,10,9}\rangle\bigr)
 \label{eq:52}.
\end{align}
The energy denominator of the intermediate state is given by
\begin{align}
E_{0-10}^n+5\lambda-3(\lambda-2)=E_{0-10}^n+2\lambda+6
 \label{eq:53}.
\end{align}
Thus, from Eqs. (\ref{eq:52}) and (\ref{eq:53}), we obtain
\begin{align}
\epsilon_1^{(\rm{II},\rm{s})}=-\frac{\Delta_{\rm{II}}^2}{4}\sum_{n}&\bigl(\langle\Psi^n_{0-10}|t_3^0;t_7^0;\phi^g_{0,2,1};\phi^g_{4,6,5};\phi^g_{8,10,9}\rangle-\langle\Psi^n_{0-10}|t_3^+;t_7^-;\phi^g_{0,2,1};\phi^g_{4,6,5};\phi^g_{8,10,9}\rangle \notag \\
&-\langle\Psi^n_{0-10}|t_3^-;t_7^+;\phi^g_{0,2,1};\phi^g_{4,6,5};\phi^g_{8,10,9}\rangle\bigr)^2/\left(E_{0-10}^n+2\lambda+6\right)
\label{eq:54}.
\end{align}

Next, we consider the lower process $\epsilon_1^{(\rm{II},\rm{t})}$ in Fig.~\ref{fig:7}(b). When $V_{5,11}$ operates on the initial state $|\phi^g_{0,2,1};\phi^g_{4,6,5};\phi^g_{8,10,9};s_{11}\rangle$, we have
\begin{align}
   V_{5,11}|\phi^g_{0,2,1};\phi^g_{4,6,5};\phi^g_{8,10,9};s_{11}\rangle=&\frac{\Delta_{\rm{II}}}{2\sqrt{3}}|s\rangle_{5}\left(|T^+\rangle_{4,6}|t^-\rangle_{11}+|T^-\rangle_{4,6}|t^+\rangle_{11}-|T^0\rangle_{4,6}|t^0\rangle_{11}\right)|\phi^g_{0,2,1};\phi^g_{8,10,9}\rangle
   \label{eq:55}.
\end{align}
If the triplet and singlet states at sites $5$ and $11$, respectively, replace each other, and site $11$ is in the triplet state, then a connected cluster $(0,1,2,8,9,10,11)$ appears in the intermediate state.
For the unperturbed Hamiltonian, $h_{0,2}+h_{2,8}+h_{8,10}$ for this cluster, and we write the eigenvalues and eigenstates as $E_{0,1,2,8,9,10,11}^n+3\lambda$ and $|\Psi^n\rangle_{0,1,2,8,9,10,11}$, respectively.
Denoting the intermediate states as $|\sigma_{4};s_{5};\sigma_{6};\Psi^n_{0,1,2,8,9,10,11}\rangle$ with $\sigma=\uparrow, \downarrow$, the matrix elements are written by 
\begin{align}
\langle&\sigma_{4};s_{5};\sigma_{6};\Psi^n_{0,1,2,8,9,10,11}|V_{5,11}|\phi^g_{0,2,1};\phi^g_{4,6,5};\phi^g_{8,10,9};s_{11}\rangle \notag \\
=&\frac{\Delta_{\rm{II}}}{2\sqrt{3}}\left(\delta_{\sigma_6,\uparrow}\delta_{\sigma_9,\uparrow}\langle\Psi^n_{0,1,2,8,9,10,11}|t^-_{11};\phi^g_{0,2,1};\phi^g_{8,10,9}\rangle+\delta_{\sigma_6,\downarrow}\delta_{\sigma_9,\downarrow}\langle\Psi^n_{0,1,2,8,9,10,11}|t^+_{11};\phi^g_{0,2,1};\phi^g_{8,10,9}\rangle \right) \notag \\
&-\frac{\Delta_{\rm{II}}}{2\sqrt{6}}\left(\delta_{\sigma_6,\uparrow}\delta_{\sigma_9,\downarrow}+\delta_{\sigma_6,\downarrow}\delta_{\sigma_9,\uparrow}\right)\langle\Psi^n_{0,1,2,8,9,10,11}|t^0_{11};\phi^g_{0,2,1};\phi^g_{8,10,9}\rangle
\label{eq:56}.
\end{align}
The energy denominator of the intermediate state is given by
\begin{align}
E_{0,1,2,8,9,10,11}^n+3\lambda-3(\lambda-2)=E_{0,1,2,8,9,10,11}^n+6
 \label{eq:57}.
\end{align}
Thus, from Eqs. (\ref{eq:56}) and (\ref{eq:57}), we obtain 
\begin{align}
\epsilon_1^{(\rm{II},\rm{t})}=&-\Delta_{\rm{II}}^2\sum_{n}\sum_{\sigma_4=\uparrow,\downarrow}\sum_{\sigma_6=\uparrow,\downarrow}\frac{{|\langle\sigma_{4};s_{5};\sigma_{6};\Psi^n_{0,1,2,8,9,10,11}|V_{5,11}|\phi^g_{0,2,1};\phi^g_{4,6,5};\phi^g_{8,10,9};s_{11}\rangle|}^2}{E_{0,1,2,8,9,10,11}^n+6} \notag \\
=&-\frac{\Delta_{\rm{II}}^2}{12}\sum_{n}\bigl({|\langle\Psi^n_{0,1,2,8,9,10,11}|t^+_{11};\phi^g_{0,2,1};\phi^g_{8,10,9}\rangle|}^2+{|\langle\Psi^n_{0,1,2,8,9,10,11}|t^-_{11};\phi^g_{0,2,1};\phi^g_{8,10,9}\rangle|}^2 \notag \\
&\;\;\;\;\;\;\;\;\;\;+{|\langle\Psi^n_{0,1,2,8,9,10,11}|t^0_{11};\phi^g_{0,2,1};\phi^g_{8,10,9}\rangle|}^2\bigr)/\left(E_{0,1,2,8,9,10,11}^n+6\right)
 \label{eq:58}.
\end{align}
Therefore, substituting Eqs. (\ref{eq:54}), (\ref{eq:58}) into Eq. (\ref{eq:37}), we obtain $\epsilon_1^{(\rm{II})}$.

\subsubsection*{A.2.3 \;Calculation of the energy $\epsilon_0^{(\rm{II})}$}

We consider the process $\epsilon_0^{(\rm{II},\rm{s})}$ in Fig.~\ref{fig:7}(c).
Because we have the same contributions from the process produced by $V_{3,11}$ and $V_{7,15}$, we consider the process using $V_{3,11}$ here. 
When $V_{3,11}$ operates on the initial state $|\phi^g_{0,2,1};s_{3};\phi^g_{4,6,5};\phi^g_{8,10,9};s_{11};\phi^g_{12,14,13}\rangle$, we have 
\begin{align}
   V_{3,11}&|\phi^g_{0,2,1};s_{3};\phi^g_{4,6,5};\phi^g_{8,10,9};s_{11};\phi^g_{12,14,13}\rangle\notag \\
   =&\frac{\Delta_{\rm{II}}}{2}\left(|t^0\rangle_3|t^0\rangle_{11}-|t^+\rangle_3|t^-\rangle_{11}-|t^-\rangle_3|t^+\rangle_{11}\right)|\phi^g_{0,2,1};\phi^g_{4,6,5};\phi^g_{8,10,9};\phi^g_{12,14,13}\rangle
   \label{eq:59}.
\end{align}
If the bond spin-pairs at sites $3$ and $11$ are in triplet states, then the connected clusters $(0,1,2,3,4,5,6)$ and $(8,9,10,11,12,13,14)$ appear in the intermediate state. 
For the unperturbed Hamiltonian, $h_{0,2}+h_{2,4}+h_{4,6}+h_{8,10}+h_{8,12}+h_{12,14}$ for these clusters, and we write the eigenvalues and eigenstates as $E_{0-6}^n+E_{8-14}^{n'}+6\lambda$ and $|\Psi^n\rangle_{0-10}|\Psi^{n'}\rangle_{8-14}$, respectively.
Therefore, the matrix elements are written by 
\begin{align}
\langle\Psi^n_{0-6};\Psi^{n'}_{8-14}|V_{3,11}|\phi^g_{0,2,1};s_{3}&;\phi^g_{4,6,5};\phi^g_{8,10,9};s_{11};\phi^g_{12,14,13}\rangle \notag\\
=&\frac{\Delta_{\rm{II}}}{2}\bigl(\langle\Psi^n_{0-6}|t_3^0;\phi^g_{0,2,1};\phi^g_{4,6,5}\rangle\langle\Psi^{n'}_{8-14}|t_{11}^0;\phi^g_{8,10,9};\phi^g_{12,14,13}\rangle \notag \\
&-\langle\Psi^n_{0-6}|t_3^+;\phi^g_{0,2,1};\phi^g_{4,6,5}\rangle\langle\Psi^{n'}_{8-14}|t_{11}^-;\phi^g_{8,10,9};\phi^g_{12,14,13}\rangle \notag \\
&-\langle\Psi^n_{0-6}|t_3^-;\phi^g_{0,2,1};\phi^g_{4,6,5}\rangle\langle\Psi^{n'}_{8-14}|t_{11}^+;\phi^g_{8,10,9};\phi^g_{12,14,13}\rangle\bigr)
 \label{eq:60}.
\end{align}
The energy denominator of the intermediate state is given by
\begin{align}
E_{0-6}^n+E_{8-14}^{n'}+6\lambda-4(\lambda-2)=E_{0-6}^n+E_{8-14}^{n'}+2\lambda+8
 \label{eq:61}.
\end{align}
Thus, from Eqs. (\ref{eq:60}) and (\ref{eq:61}), we obtain
\begin{align}
\epsilon_0^{(\rm{II},\rm{s})}=-\frac{\Delta_{\rm{II}}^2}{4}&\sum_{n,n'}\Bigl({|\langle\Psi^n_{0-6}|t_3^0;\phi^g_{0,2,1};\phi^g_{4,6,5}\rangle\langle\Psi^{n'}_{8-14}|t_{11}^0;\phi^g_{8,10,9};\phi^g_{12,14,13}\rangle|}^2 \notag \\
&+{|\langle\Psi^n_{0-6}|t_3^+;\phi^g_{0,2,1};\phi^g_{4,6,5}\rangle\langle\Psi^{n'}_{8-14}|t_{11}^-;\phi^g_{8,10,9};\phi^g_{12,14,13}\rangle|}^2\notag \\
&+{|\langle\Psi^n_{0-6}|t_3^-;\phi^g_{0,2,1};\phi^g_{4,6,5}\rangle\langle\Psi^{n'}_{8-14}|t_{11}^+;\phi^g_{8,10,9};\phi^g_{12,14,13}\rangle|}^2\Bigr)/(E_{0-6}^n+E_{8-14}^{n'}+2\lambda+8)
\label{eq:62},
\end{align}
where we use the fact that the cross term of the square of Eq. (\ref{eq:60}) is zero.
Therefore, substituting Eq. (\ref{eq:62}) into Eq. (\ref{eq:38}), we obtain $\epsilon_0^{(\rm{II})}$.

\subsection*{A.3 \;Calculation process for the pair-hopping amplitude $t^{(\rm{I})}$}

\begin{figure}[!htbp]
\begin{center}
\includegraphics[width=.70\linewidth]{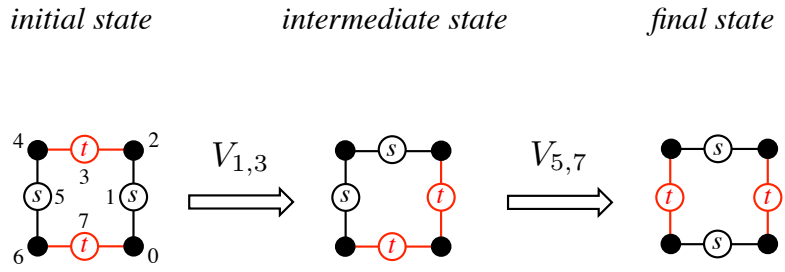}
\end{center}
\caption{(Color online) Second-order perturbation process for the pair hopping of dimers when we use the coupling $\Delta_{\rm{I}}$.
In addition to the process shown in this figure, we have three other processes; (initial state)$\stackrel{V_{1,7}}{\to}$(intermediate state)$\stackrel{V_{3,5}}{\to}$(final state), (initial state)$\stackrel{V_{3,5}}{\to}$(intermediate state)$\stackrel{V_{1,7}}{\to}$(final state), and (initial state)$\stackrel{V_{5,7}}{\to}$(intermediate state)$\stackrel{V_{1,3}}{\to}$(final state). All four processes have the same pair-hopping amplitudes.}
\label{fig:9}
\end{figure}

In Fig.~\ref{fig:9}, we show a possible second-order perturbation process for the pair hopping of dimers when we use the coupling $\Delta_{\rm{I}}$.
When $V_{1,3}$ operates on the initial state $|s_{1};\phi^g_{2,4,3};s_{5};\phi^g_{0,6,7}\rangle$, the triplet and singlet states at sites $3$ and $1$, respectively, are replaced by each other, and we have a connected cluster $(0,1,2,6,7)$ in the intermediate state. Furthermore, when $V_{5,7}$ operates on the intermediate state, the triplet and singlet states at sites $7$ and $5$, respectively, are replaced by each other, and we have the final state $|\phi^g_{0,2,1};s_{3};\phi^g_{4,6,5};s_{7}\rangle$. If we write the process for the pair hopping of dimers shown in Fig.~\ref{fig:9} as (initial state)$\stackrel{V_{1,3}}{\to}$(intermediate state)$\stackrel{V_{5,7}}{\to}$(final state), we have three other processes; (initial state)$\stackrel{V_{1,7}}{\to}$(intermediate state)$\stackrel{V_{3,5}}{\to}$(final state), (initial state)$\stackrel{V_{3,5}}{\to}$(intermediate state)$\stackrel{V_{1,7}}{\to}$(final state), and (initial state)$\stackrel{V_{5,7}}{\to}$(intermediate state)$\stackrel{V_{1,3}}{\to}$(final state).
All four processes have the same pair-hopping amplitudes, and summing these four amplitudes yields
\begin{align}
t^{(\rm{I})}=1.06026786\Delta_{\rm{I}}^2
\label{eq:63},
\end{align}
which was calculated in our previous paper\cite{Hirose2016}.

Here, we mention the reason why Eq. (\ref{eq:63}) does not depend on $\lambda$. 
Because there are two dimers in the intermediate state, as shown in Fig.~\ref{fig:9}, we can write the eigenvalues in the intermediate state as $E^n_{0,1,2,6,7}+2\lambda$. Therefore, the energy denominator of the intermediate state can be obtained by $E^n_{0,1,2,6,7}+2\lambda-2(\lambda-2)=E^n_{0,1,2,6,7}+4$ or Eq. (26) of Ref. 27, which indicates that there is no $\lambda$ dependence. The dependence on $\lambda$ originates from the presence or absence of $\lambda$ in the energy denominator of the intermediate state, i.e., the difference between the number of dimers in the intermediate state and that in the initial (final) state in the perturbation process. In the case of $t^{(\rm{I})}$, because the number of dimers is two in both the initial (final) and intermediate states, there is no $\lambda$ dependence. On the other hand, in the case of $t^{(\rm{II})}$, there is $\lambda$ dependence because the number of dimers is different in the initial (final) and intermediate states, as will be discussed later.

\subsection*{A.4 \;Calculation process for the pair-hopping amplitude $t^{(\rm{II})}$}

\begin{figure}[!htbp]
\begin{center}
\includegraphics[width=.99\linewidth]{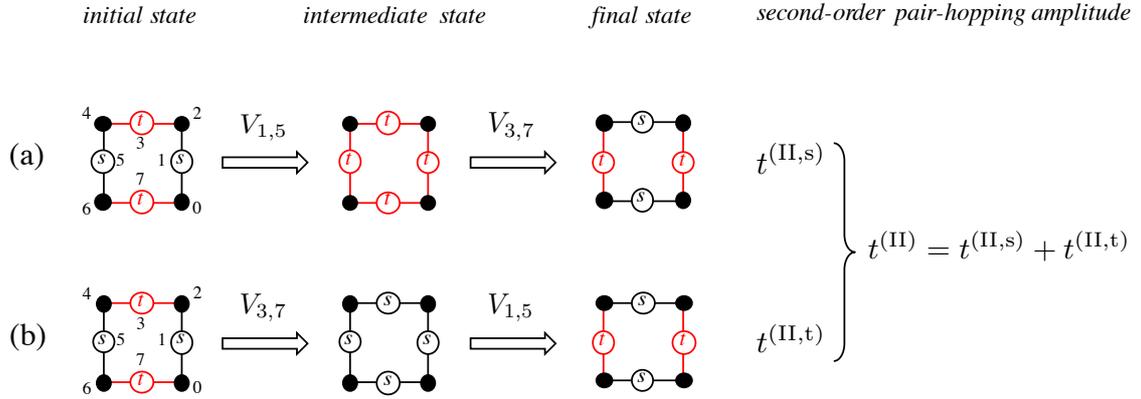}
\end{center}
\caption{(Color online) Second-order perturbation processes for the pair hopping of dimers when we use the coupling $\Delta_{\rm{II}}$.
There are two kinds of processes, (a) and (b).}
\label{fig:10}
\end{figure}

In Fig.~\ref{fig:10}, we show a possible second-order perturbation process for the pair hopping of dimers when we use the coupling $\Delta_{\rm{II}}$.
In this case, there are two kinds of processes, as shown in Figs.~\ref{fig:10}(a) and \ref{fig:10}(b). 
In Fig.~\ref{fig:10}(a), when we use the operator $V_{1,5}$, the singlet states at sites $1$ and $5$ turn into triplet states, and all the bond spin-pairs at sites $1$, $3$, $5$, and $7$ change to triplet states in the intermediate state. 
Furthermore, when $V_{3,7}$ operates on the intermediate state and the triplet states at sites $3$ and $7$ turn into singlet states, the pair hopping of dimers occurs.
In Fig.~\ref{fig:10}(b), when we use the operator $V_{3,7}$, the triplet states at sites $3$ and $7$ turn into singlet states, and all the bond spin-pairs at sites $1$, $3$, $5$, and $7$ change to singlet states in the intermediate state. 
Furthermore, when $V_{1,5}$ operates on the intermediate state and the singlet states at sites $1$ and $5$, respectively, turning them into triplet states, the pair hopping of dimers occurs.
Defining the second-order pair-hopping amplitudes in the processes in Figs.~\ref{fig:10}(a) and \ref{fig:10}(b) as $t^{(\rm{II},s)}$ and $t^{(\rm{II},t)}$, respectively, we can write 
\begin{align}
t^{(\rm{II})}=t^{(\rm{II},s)}+t^{(\rm{II},t)}
\label{eq:64}.
\end{align}

\begin{figure}[h]
\begin{center}
\includegraphics[width=.80\linewidth]{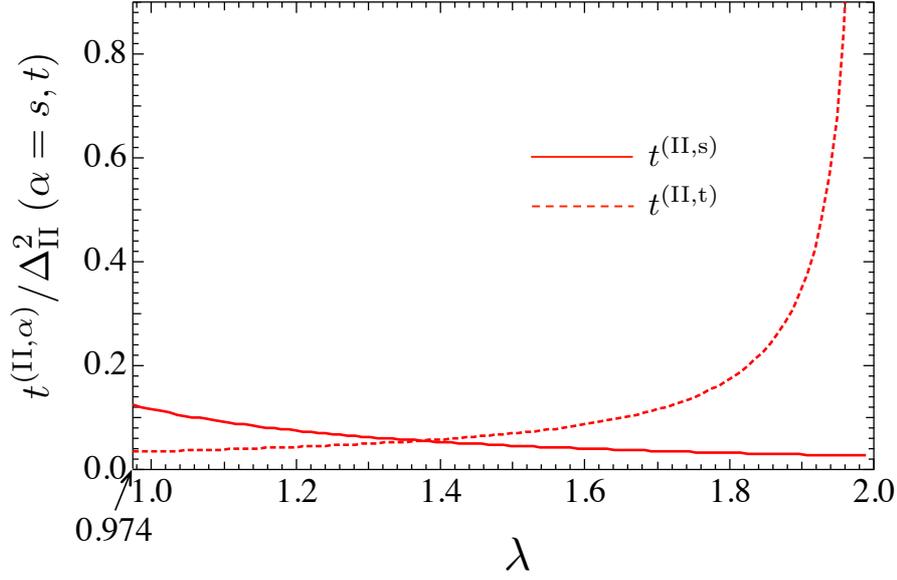}
\end{center}
\caption{(Color online) Calculation results for the dependence of $t_2^{\rm{(II,s)}}$ and $t^{\rm{(II,t)}}$ on $\lambda$.}
\label{fig:11}
\end{figure}

In Fig.~\ref{fig:11}, we show the numerical calculation results for the dependence of $t^{(\rm{II},s)}$ and $t^{(\rm{II},t)}$ on $\lambda$. 
These results show that the pair-hopping amplitude $t^{(\rm{II},s)}$ gradually decreases for $\lambda\to2$. On the other hand, $t^{(\rm{II},t)}$ increases and diverges to $+\infty$ at $\lambda=2$, which produces the divergence of $t^{(\rm{II})}\to+\infty$ at $\lambda=2$, as will be discussed later.


\subsubsection*{A.4.1 \;Calculation of the pair-hopping amplitude $t^{(\rm{II,s})}$}

We consider the process $t^{(\rm{II},s)}$ in Fig.~\ref{fig:10}(a).
When $V_{1,5}$ operates on the initial state $|s_{1};\phi^g_{2,4,3};s_{5};\phi^g_{0,6,7}\rangle$, we have Eq. (\ref{eq:39}).
Therefore, the matrix elements are written by Eq. (\ref{eq:40}) and the energy denominator of the intermediate state is given by Eq. (\ref{eq:41}).

Next, when $V_{3,7}$ operates on the final state $|\phi^g_{0,2,1};s_{3};\phi^g_{4,6,5};s_{7}\rangle$, we have 
\begin{equation}
   V_{3,7}|\phi^g_{0,2,1};s_{3};\phi^g_{4,6,5};s_{7}\rangle=\frac{\Delta_{\rm{II}}}{2}\left(|t^0\rangle_3|t^0\rangle_{7}-|t^+\rangle_3|t^-\rangle_{7}-|t^-\rangle_3|t^+\rangle_{7}\right)|\phi^g_{0,2,1};\phi^g_{4,6,5}\rangle
   \label{eq:65}.
\end{equation}
Thus, the matrix elements between the final state and an intermediate state $|\Psi^n\rangle_{0-7}$ are written by
\begin{align}
\langle&\Psi^n_{0-7}|V_{3,7}|\phi^g_{0,2,1};s_{3};\phi^g_{4,6,5};s_{7}\rangle \notag\\
=&\frac{\Delta_{\rm{II}}}{2}\left(\langle\Psi^n_{0-7}|t_3^0;t_7^0;\phi^g_{0,2,1};\phi^g_{4,6,5}\rangle-\langle\Psi^n_{0-7}|t_3^+;t_7^-;\phi^g_{0,2,1};\phi^g_{4,6,5}\rangle-\langle\Psi^n_{0-7}|t_3^-;t_7^+;\phi^g_{0,2,1};\phi^g_{4,6,5}\rangle\right)
 \label{eq:66}.
\end{align}
Thus, from Eqs. (\ref{eq:40}), (\ref{eq:41}), and (\ref{eq:66}), we obtain
\begin{align}
t^{(\rm{II},\rm{s})}=&\Delta_{\rm{II}}^2\sum_{n}\frac{\langle\phi^g_{0,2,1};s_{3};\phi^g_{4,6,5};s_{7}|V_{3,7}|\Psi^n_{0-7}\rangle\langle\Psi^n_{0-7}|V_{1,5}|s_{1};\phi^g_{2,4,3};s_{5};\phi^g_{0,6,7}\rangle}{E_{0-7}^n+2\lambda+4}, \notag \\
			 =&\Delta_{\rm{II}}^2\sum_{n}\frac{\langle\Psi^n_{0-7}|V_{3,7}|\phi^g_{0,2,1};s_{3};\phi^g_{4,6,5};s_{7}\rangle\langle\Psi^n_{0-7}|V_{1,5}|s_{1};\phi^g_{2,4,3};s_{5};\phi^g_{0,6,7}\rangle}{E_{0-7}^n+2\lambda+4}, \notag \\
			 =&\frac{\Delta_{\rm{II}}^2}{4}\sum_{n}\bigl(\langle\Psi^n_{0-7}|t_3^0;t_7^0;\phi^g_{0,2,1};\phi^g_{4,6,5}\rangle-\langle\Psi^n_{0-7}|t_3^+;t_7^-;\phi^g_{0,2,1};\phi^g_{4,6,5}\rangle \notag \\
			 &-\langle\Psi^n_{0-7}|t_3^-;t_7^+;\phi^g_{0,2,1};\phi^g_{4,6,5}\rangle\bigr)
			 \times\bigl(\langle\Psi^n_{0-7}|t_1^+;t_5^-;\phi^g_{2,4,3};\phi^g_{0,6,7}\rangle+\langle\Psi^n_{0-7}|t_1^-;t_5^+;\phi^g_{2,4,3};\phi^g_{0,6,7}\rangle \notag \\
			 &-\langle\Psi^n_{0-7}|t_1^0;t_5^0;\phi^g_{2,4,3};\phi^g_{0,6,7}\rangle\bigr)/\left(E_{0-7}^n+2\lambda+4\right)
\label{eq:67},
\end{align}
where Eq. (\ref{eq:67}) depends on $\lambda$ because the numbers of dimers in the intermediate and initial (final) states, four and two dimers, respectively, are different, and the energy denominator of the intermediate state is written by $E_{0-7}^n+2\lambda+4$.

\subsubsection*{A.4.2 \;Calculation of the pair-hopping amplitude $t^{(\rm{II,t})}$}

We consider the process $t^{(\rm{II},t)}$ in Fig.~\ref{fig:10}(b).
From Eq. (\ref{eq:43}), when $V_{3,7}$ operates on the initial state $|s_{1};\phi^g_{2,4,3};s_{5};\phi^g_{0,6,7}\rangle$, we have 
\begin{align}
V_{3,7}&|s_{1};\phi^g_{2,4,3};s_{5};\phi^g_{0,6,7}\rangle \notag \\
=&(\text{the states where sites $3$ and $7$ are in triplet states}) \notag \\
&-\frac{\Delta_{\rm{II}}}{6}\left(|\downarrow_0;\uparrow_2;\uparrow_4;\downarrow_6\rangle+|\uparrow_0;\downarrow_2;\downarrow_4;\uparrow_6\rangle\right)|s_1;s_3;s_5;s_7\rangle \notag \\
&-\frac{\Delta_{\rm{II}}}{12}\left(|\downarrow_0;\uparrow_2;\downarrow_4;\uparrow_6\rangle+|\uparrow_0;\uparrow_2;\downarrow_4;\downarrow_6\rangle+|\downarrow_0;\downarrow_2;\uparrow_4;\uparrow_6\rangle+|\uparrow_0;\downarrow_2;\uparrow_4;\downarrow_6\rangle\right)|s_1;s_3;s_5;s_7\rangle
\label{eq:68}.
\end{align}
The energy denominator of the intermediate state is given by Eq. (\ref{eq:45}).

Next, when $V_{1,5}$ operates on the final state $|\phi^g_{0,2,1};s_{3};\phi^g_{4,6,5};s_{7}\rangle$, we have
\begin{align}
V_{3,7}&|\phi^g_{0,2,1};s_{3};\phi^g_{4,6,5};s_{7}\rangle \notag \\
=&(\text{the states where sites $1$ and $5$ are in triplet states}) \notag \\
&-\frac{\Delta_{\rm{II}}}{6}\left(|\downarrow_0;\downarrow_2;\uparrow_4;\uparrow_6\rangle+|\uparrow_0;\uparrow_2;\downarrow_4;\downarrow_6\rangle\right)|s_1;s_3;s_5;s_7\rangle \notag \\
&-\frac{\Delta_{\rm{II}}}{12}\left(|\downarrow_0;\uparrow_2;\uparrow_4;\downarrow_6\rangle+|\uparrow_0;\downarrow_2;\uparrow_4;\downarrow_6\rangle+|\downarrow_0;\uparrow_2;\downarrow_4;\uparrow_6\rangle+|\uparrow_0;\downarrow_2;\downarrow_4;\uparrow_6\rangle\right)|s_1;s_3;s_5;s_7\rangle
\label{eq:69}.
\end{align}
Thus, from Eqs. (\ref{eq:45}), (\ref{eq:68}), and (\ref{eq:69}), we can write
\begin{align}
t^{(\rm{II},\rm{t})} 
=&\frac{\Delta_{\rm{II}}^2}{4-2\lambda}\Bigl(\langle\phi^g_{0,2,1};s_{3};\phi^g_{4,6,5};s_{7}|V_{1,5}|\downarrow_0;\uparrow_2;\uparrow_4;\downarrow_6;s_1;s_3;s_5;s_7\rangle \notag \\
&\times\langle\downarrow_0;\uparrow_2;\uparrow_4;\downarrow_6;s_1;s_3;s_5;s_7|V_{3,7}|s_{1};\phi^g_{2,4,3};s_{5};\phi^g_{0,6,7}\rangle \notag \\
&+\langle\phi^g_{0,2,1};s_{3};\phi^g_{4,6,5};s_{7}|V_{1,5}|\uparrow_0;\downarrow_2;\downarrow_4;\uparrow_6;s_1;s_3;s_5;s_7\rangle \notag\\
&\times\langle\uparrow_0;\downarrow_2;\downarrow_4;\uparrow_6;s_1;s_3;s_5;s_7|V_{3,7}|s_{1};\phi^g_{2,4,3};s_{5};\phi^g_{0,6,7}\rangle \notag \\
&+\langle\phi^g_{0,2,1};s_{3};\phi^g_{4,6,5};s_{7}|V_{1,5}|\downarrow_0;\uparrow_2;\downarrow_4;\uparrow_6;s_1;s_3;s_5;s_7\rangle \notag\\
&\times\langle\downarrow_0;\uparrow_2;\downarrow_4;\uparrow_6;s_1;s_3;s_5;s_7|V_{3,7}|s_{1};\phi^g_{2,4,3};s_{5};\phi^g_{0,6,7}\rangle \notag \\
&+\langle\phi^g_{0,2,1};s_{3};\phi^g_{4,6,5};s_{7}|V_{1,5}|\uparrow_0;\uparrow_2;\downarrow_4;\downarrow_6;s_1;s_3;s_5;s_7\rangle \notag\\
&\times\langle\uparrow_0;\uparrow_2;\downarrow_4;\downarrow_6;s_1;s_3;s_5;s_7|V_{3,7}|s_{1};\phi^g_{2,4,3};s_{5};\phi^g_{0,6,7}\rangle \notag \\
&+\langle\phi^g_{0,2,1};s_{3};\phi^g_{4,6,5};s_{7}|V_{1,5}|\downarrow_0;\downarrow_2;\uparrow_4;\uparrow_6;s_1;s_3;s_5;s_7\rangle \notag\\
&\times\langle\downarrow_0;\downarrow_2;\uparrow_4;\uparrow_6;s_1;s_3;s_5;s_7|V_{3,7}|s_{1};\phi^g_{2,4,3};s_{5};\phi^g_{0,6,7}\rangle \notag \\
&+\langle\phi^g_{0,2,1};s_{3};\phi^g_{4,6,5};s_{7}|V_{1,5}|\uparrow_0;\downarrow_2;\uparrow_4;\downarrow_6;s_1;s_3;s_5;s_7\rangle \notag\\
&\times\langle\uparrow_0;\downarrow_2;\uparrow_4;\downarrow_6;s_1;s_3;s_5;s_7|V_{3,7}|s_{1};\phi^g_{2,4,3};s_{5};\phi^g_{0,6,7}\rangle\Bigr)
\label{eq:70}.
\end{align}
Furthermore, from Eqs. (\ref{eq:68}) and (\ref{eq:69}), we can write
\begin{align}
-\frac{\Delta_{\rm{II}}}{12}=&\langle\phi^g_{0,2,1};s_{3};\phi^g_{4,6,5};s_{7}|V_{1,5}|\downarrow_0;\uparrow_2;\uparrow_4;\downarrow_6;s_1;s_3;s_5;s_7\rangle	\notag \\
		  =&\langle\phi^g_{0,2,1};s_{3};\phi^g_{4,6,5};s_{7}|V_{1,5}|\uparrow_0;\downarrow_2;\downarrow_4;\uparrow_6;s_1;s_3;s_5;s_7\rangle \notag \\
		 =&\langle\phi^g_{0,2,1};s_{3};\phi^g_{4,6,5};s_{7}|V_{1,5}|\downarrow_0;\uparrow_2;\downarrow_4;\uparrow_6;s_1;s_3;s_5;s_7\rangle\notag \\
		 =&\langle\downarrow_0;\uparrow_2;\downarrow_4;\uparrow_6;s_1;s_3;s_5;s_7|V_{3,7}|s_{1};\phi^g_{2,4,3};s_{5};\phi^g_{0,6,7}\rangle\notag \\
		 =&\langle\uparrow_0;\uparrow_2;\downarrow_4;\downarrow_6;s_1;s_3;s_5;s_7|V_{3,7}|s_{1};\phi^g_{2,4,3};s_{5};\phi^g_{0,6,7}\rangle \notag \\
		 =&\langle\downarrow_0;\downarrow_2;\uparrow_4;\uparrow_6;s_1;s_3;s_5;s_7|V_{3,7}|s_{1};\phi^g_{2,4,3};s_{5};\phi^g_{0,6,7}\rangle \notag \\
		=&\langle\phi^g_{0,2,1};s_{3};\phi^g_{4,6,5};s_{7}|V_{1,5}|\uparrow_0;\downarrow_2;\uparrow_4;\downarrow_6;s_1;s_3;s_5;s_7\rangle \notag \\
		=&\langle\uparrow_0;\downarrow_2;\uparrow_4;\downarrow_6;s_1;s_3;s_5;s_7|V_{3,7}|s_{1};\phi^g_{2,4,3};s_{5};\phi^g_{0,6,7}\rangle,
\label{eq:71}
\end{align}
and
\begin{align}
-\frac{\Delta_{\rm{II}}}{6}=&\langle\downarrow_0;\uparrow_2;\uparrow_4;\downarrow_6;s_1;s_3;s_5;s_7|V_{3,7}|s_{1};\phi^g_{2,4,3};s_{5};\phi^g_{0,6,7}\rangle \notag \\
		=&\langle\uparrow_0;\downarrow_2;\downarrow_4;\uparrow_6;s_1;s_3;s_5;s_7|V_{3,7}|s_{1};\phi^g_{2,4,3};s_{5};\phi^g_{0,6,7}\rangle \notag \\
		=&\langle\phi^g_{0,2,1};s_{3};\phi^g_{4,6,5};s_{7}|V_{1,5}|\uparrow_0;\uparrow_2;\downarrow_4;\downarrow_6;s_1;s_3;s_5;s_7\rangle \notag \\
		=&\langle\phi^g_{0,2,1};s_{3};\phi^g_{4,6,5};s_{7}|V_{1,5}|\downarrow_0;\downarrow_2;\uparrow_4;\uparrow_6;s_1;s_3;s_5;s_7\rangle
\label{eq:72}.
\end{align}
Substituting Eqs. (\ref{eq:71}) and (\ref{eq:72}) into Eq. (\ref{eq:70}), we obtain
\begin{align}
t^{(\rm{II},\rm{t})} =\frac{5}{144(2-\lambda)}\Delta_{\rm{II}}^2
\label{eq:73},
\end{align}
where $t^{(\rm{II},\rm{t})}$ depends on $\lambda$ because the numbers of dimers in the intermediate and initial (final) states, zero and two dimers, respectively, are different, and the energy denominator of the intermediate state is written by $4-2\lambda$.
Furthermore, we find that the denominator of the right-hand side of Eq. (\ref{eq:73}) produces the divergence of $t^{(\rm{II})}\to+\infty$ at $\lambda=2$.
Then, substituting Eqs. (\ref{eq:67}) and (\ref{eq:73}) into Eq. (\ref{eq:64}), we obtain $t^{(\rm{II})}$.

Note that from Eqs. (\ref{eq:46}) and (\ref{eq:73}), we obtain $v\approx\frac{\Delta_{\rm{II}}^2}{24(\lambda-2)}$ and $t\approx\frac{5\Delta_{\rm{II}}^2}{144(2-\lambda)}$ for $\lambda\to2$. Therefore, $v/t$ converges to
\begin{align}
v/t=\frac{\Delta_{\rm{II}}^2}{24(\lambda-2)}\bigg/\frac{5\Delta_{\rm{II}}^2}{144(2-\lambda)}=-1.2
\label{eq:74}
\end{align}
for $\lambda\to2$, as shown in Fig.~\ref{fig:5}.


\section*{Acknowledgment}
This work was supported by JSPS KAKENHI Grant Numbers JP17J05190 and JP17K05519.


\begin{thebibliography}{99}
\bibitem{Anderson1973} P.~W.~Anderson, Math. Res. Bull. \textbf{8}, 153 (1973).
\bibitem{Rokhsar1988} D. S. Rokhsar and S. A. Kivelson, Phys. Rev. Lett. \textbf{61}, 2376 (1988).
\bibitem{square_qdm_1996} P. W. Leung, K. C. Chiu, and K. J. Runge, Phys. Rev. B \textbf{54}, 12938 (1996).
\bibitem{square_qdm_2006} O. F. Sylju\r{a}sen, Phys. Rev. B \textbf{73}, 245105 (2006).
\bibitem{square_qdm_2008} A. Ralko, D. Poilblanc, and R. Moessner, Phys. Rev. Lett. \textbf{100}, 037201 (2008).
\bibitem{square_qdm_2014} D. Banerjee, M. B$\ddot{\rm{o}}$gli, C. P. Hofmann, F.-J. Jiang, P. Widmer, and U.-J. Wiese, Phys. Rev. B \textbf{90}, 245143 (2014).
\bibitem{square_qdm_2016} D. Banerjee, M. B$\ddot{\rm{o}}$gli, C. P. Hofmann, F.-J. Jiang, P. Widmer, and U.-J. Wiese, Phys. Rev. B \textbf{94}, 115120 (2016).
\bibitem{Moessner2008} R. Moessner and K. S. Raman, arXiv:cond-mat/0809.3051v1.
\bibitem{book} \textit{Introduction to Frustrated Magnetism}, ed. C. Lacroix, P. Mendels, and F. Mila, Springer Series in Solid-State Sciences, (Springer, Heidelberg, 2011) Vol. 164.
\bibitem{Moessner2001} R. Moessner and S. L. Sondhi, Phys. Rev. Lett. \textbf{86}, 1881 (2001).
\bibitem{Moessner2002} R. Moessner and S. L. Sondhi, Prog. Theor. Phys. Suppl. \textbf{145}, 37 (2002).
\bibitem{Moessner2005} A. Ralko, M. Ferrero, F. Becca, D. Ivanov, and F. Mila, Phys. Rev. B \textbf{71}, 224109 (2005).
\bibitem{Huse2003} D. A. Huse, W. Krauth, R. Moessner, and S. L. Sondhi, Phys. Rev. Lett. \textbf{91}, 167004 (2003).
\bibitem{Moessner2003} R. Moessner and S. L. Sondhi, Phys. Rev. B \textbf{68}, 184512 (2003).
\bibitem{Hermele2004} M. Hermele, M. P. A. Fisher, and L. Balents, Phys. Rev. B \textbf{69}, 064404 (2004).
\bibitem{Fujimoto} S. Fujimoto, Phys. Rev. B \textbf{72}, 024429 (2005).
\bibitem{Yan2011} S. Yan, D. A. Huse, and S. R. White, Science \textbf{332}, 1173 (2011).
\bibitem{Depenbrock2012} S. Depenbrock, I. P. McCulloch, and U. Schollw\"ock, Phys. Rev. Lett. \textbf{109}, 067201 (2012).
\bibitem{herbertsmithite1} P. Mendels, F. Bert, M. A. de. Vries, A. Olariu, A. Harrison, F. Duc, J. C. Trombe, J. S. Lord, A. Amato, and C. Baines, Phys. Rev. Lett. \textbf{98}, 077204 (2007).
\bibitem{herbertsmithite2} J. S. Helton, K. Matan, M. P. Shores, E. A. Nytko, B. M. Bartlett, Y. Yoshida, Y. Takano, A. Suslov, Y. Qiu, J.-H. Chung, D. G. Nocera, and Y. S. Lee, Phys. Rev. Lett. \textbf{98}, 107204 (2007).
\bibitem{herbertsmithite3} T.-H. Han, J. S. Helton, S. Chu, D. G. Nocera, J. A. R.-Rivera, C. Broholm, and Y. S. Lee, Nature \textbf{492}, 406 (2012).
\bibitem{Kawamura1} H. Kawamura, K. Watanabe, and T. Shimokawa, J. Phys. Soc. Jpn. \textbf{83}, 103704 (2014).
\bibitem{Kawamura2} T. Shimokawa, K. Watanabe, and H. Kawamura, Phys. Rev. B \textbf{92}, 134407 (2015).
\bibitem{Kitaev1} A. Kitaev, Ann. Phys. \textbf{321}, 2 (2006).
\bibitem{Kitaev2} G. Baskaran, S. Mandal, and R. Shankar, Phys. Rev. Lett. \textbf{98}, 247201 (2007).
\bibitem{Kitaev3} S. K. Choi, R. Coldea, A. N. Kolmogorov, T. Lancaster, I. I. Mazin, S. J. Blundell, P. G. Radaelli, Y. Singh, P. Gegenwart, K. R. Choi, S.-W. Cheong, P. J. Baker, C. Stock, and J. Taylor, Phys. Rev. Lett. \textbf{108}, 127204 (2012).
\bibitem{Hirose2016} Y. Hirose, A. Oguchi, and Y. Fukumoto, J. Phys. Soc. Jpn. \textbf{85}, 094002 (2016).
\bibitem{Bloch2008} I. Bloch, J. Dalibard, and W. Zwerger, Rev. Mod. Phys.  \textbf{80}, 885 (2008).
\bibitem{raman1} P. Lemmens, M. Grove, M. Fischer, and G. G$\ddot{\rm{u}}$ntherodt, Phys. Rev. Lett. \textbf{85}, 2605 (2000).
\bibitem{Hirose2017} Y. Hirose, A. Oguchi, and Y. Fukumoto, J. Phys. Soc. Jpn. \textbf{86}, 014002 (2017).
\bibitem{Takano1996} K. Takano, K. Kubo, and H. Sakamoto, J. Phys.: Condens. Matter \textbf{8}, 6405 (1996).
\bibitem{Morita2016} K. Morita and N. Shibata, J. Phys. Soc. Jpn. \textbf{85}, 033705 (2016).
\bibitem{raman2} N. Kunisada and Y. Fukumoto, Prog. Theor. Exp. Phys, 041I01 (2014).
\end{thebibliography}
\end{document}